\documentclass[preprint,aps,prd,10pt,nofootinbib,superscriptaddress,preprintnumbers,titlepage,amssymb,amsmath]{revtex4-2}

\usepackage{graphicx}
\usepackage{amssymb,amsfonts,amsmath}
\usepackage{hyperref}
\usepackage{nicefrac}
\usepackage[utf8]{inputenc}
\usepackage[T1]{fontenc}
\usepackage{xcolor}

\graphicspath{{plots/}}

\newcommand{\chiPT}{$\chi$PT}
\renewcommand{\vec}[1]{\mathbf{#1}}
\renewcommand{\l}{\left}
\renewcommand{\r}{\right}
\newcommand{\g}[1]{\gamma_{#1}} % gamma matrices (covariant index)
\newcommand{\trace}[1]{\mathrm{tr}\left[ #1 \right]} % defines the trace operator
\newcommand{\tr}{\mathrm{tr}}
\newcommand{\<}{\langle}
\renewcommand{\>}{\rangle}

\newcommand{\gev}{\,\mathrm{GeV}}

\newcommand{\mev}{\,\mathrm{MeV}}
\newcommand{\fm}{\,\mathrm{fm}}

\newcommand{\SU}[1]{\mathrm{SU}\l(#1\r)}

\newcommand{\phys}{\mathrm{phys}}
\newcommand{\stat}[1]{\mathrm{stat}}
\newcommand{\sys}[1]{\mathrm{sys}}
\newcommand{\total}[1]{\mathrm{total}}
\newcommand{\staterr}[1]{(#1)_\mathrm{stat}}
\newcommand{\syserr}[1]{(#1)_\mathrm{sys}}
\newcommand{\totalerr}[1]{[#1]_\mathrm{total}}
\newcommand{\tex}{t_\mathrm{ex}}
\newcommand{\tins}{t_\mathrm{ins}}

\newcommand{\tsep}{t_\mathrm{sep}}
\newcommand{\tsepmin}{t_\mathrm{sep}^\mathrm{min}}
\newcommand{\tsepmax}{t_\mathrm{sep}^\mathrm{max}}
\newcommand{\tseplo}{t_\mathrm{sep}^\mathrm{lo}}
\newcommand{\tsephi}{t_\mathrm{sep}^\mathrm{hi}}
\newcommand{\Mpi}{M_\pi}
\newcommand{\MK}{M_\mathrm{K}}

\newcommand{\rsqr}[3]{\langle r^2_{#1} \rangle_{#2}^{#3}}
\newcommand{\rsqrpisinglet}{\rsqr{S}{\pi}{0}}
\newcommand{\rsqrpioctet}{\rsqr{S}{\pi}{8}}
\newcommand{\rsqrpi}{\rsqr{S}{\pi}{l}}
\newcommand{\FF}[4]{F_{#1}^{#2,#3}(#4)}
\newcommand{\FFpisinglet}[1]{\FF{S}{\pi}{0}{#1}}
\newcommand{\FFpioctet}[1]{\FF{S}{\pi}{8}{#1}}
\newcommand{\FFpi}[1]{\FF{S}{\pi}{l}{#1}}

\newcommand{\mfp}{\mathfrak{p}}
\newcommand{\mfpi}{\mathfrak{p}_i}
\newcommand{\mfq}{\mathfrak{q}}
\newcommand{\mfpf}{\mathfrak{p}_f}
\newcommand{\vp}{\mathbf{p}}
\newcommand{\vpi}{\mathbf{p}_i}
\newcommand{\vq}{\mathbf{q}}
\newcommand{\vpf}{\mathbf{p}_f}
\newcommand{\vx}{\mathbf{x}}
\newcommand{\vxi}{\mathbf{x}_i}
\newcommand{\vxop}{\mathbf{x}_{op}}
\newcommand{\vxf}{\mathbf{x}_f}
\newcommand{\Ctwopt}[2]{C_{#1}^\mathrm{2pt}(#2)}
\newcommand{\CtwoptHP}[2]{C_{#1}^\mathrm{2pt,HP}(#2)}
\newcommand{\CtwoptLP}[2]{C_{#1}^\mathrm{2pt,LP}(#2)}
\newcommand{\Cthreept}[4]{C_{#1\mathcal{#2}^{#3}#1}^\mathrm{3pt}(#4)}
\newcommand{\CthreeptHP}[4]{C_{#1\mathcal{#2}^{#3}#1}^\mathrm{3pt,HP}(#4)}
\newcommand{\CthreeptLP}[4]{C_{#1\mathcal{#2}^{#3}#1}^\mathrm{3pt,LP}(#4)}
\newcommand{\Cthreeptdisc}[4]{C_{#1\mathcal{#2}^{#3}#1}^{\mathrm{(2+1)pt}}(#4)}
\newcommand{\Cvev}[4]{C_{#1\mathcal{#2}^{#3}#1}^{\mathrm{vev}}(#4)}
\newcommand\mc[1]{\multicolumn{1}{c}{#1}}

\newcommand\PRISMA{\affiliation{PRISMA$^+$~Cluster~of~Excellence and Institut~f\"ur~Kernphysik, Johannes~Gutenberg-Universität~Mainz, 55099~Mainz, Germany}}
\begin{document}

 \title{\textbf{The Scalar Size of the Pion from Lattice QCD}}

 \author{Konstantin~Ottnad}  \email{kottnad@uni-mainz.de}  \PRISMA
 \author{Georg von Hippel}   \PRISMA

 \preprint{MITP-25-024}

 \date{\today}

 \begin{abstract}
  We present a lattice QCD calculation of the pion scalar form factor and associated radii with fully controlled systematics. Lattice results are computed on a large set of 17 gauge ensembles with $N_f=2+1$ Wilson Clover-improved sea quarks. These ensembles cover four values of the lattice spacing between $a=0.049\mathrm{fm}$ and $a=0.086\mathrm{fm}$, a pion mass range of $130 - 350\mathrm{MeV}$ and various physical volumes. A precise determination of the notorious quark-disconnected contributions facilitates an unprecedented momentum resolution for the form factor, particularly on large and fine ensembles in the vicinity of physical quark mass. A large range of source-sink separations $1.0\fm \lesssim \tsep \lesssim 3.25\fm$ is used to reliably extract the relevant ground state matrix elements at vanishing and non-vanishing momentum transfer. This allows us for the first time to obtain the scalar radii from a $z$-expansion parametrization of the $Q^2$-dependence of the resulting form factors rather than a simple, linear approximation at small momentum transfer. The physical extrapolation for the radii is carried out using three-flavor NLO chiral perturbation theory to parametrize the quark mass dependence in terms of three low-energy constants, including the first lattice determination of $L_4^r$. Systematic uncertainties on the final results related to the ground state extraction, form factor parametrization as well as the physical extrapolation are accounted for via model averages based on the Akaike Information Criterion.

Our physical results for the pion singlet and octet scalar radii are $\rsqrpisinglet = 0.550 \staterr{32} \syserr{39} \totalerr{50} \fm^2$ and $\rsqrpioctet = 0.443 \staterr{16} \syserr{45} \totalerr{47} \fm^2$, respectively. The values of the relevant low energy constants are given by $f_0 = 116.5\staterr{5.5}\syserr{13.7}\totalerr{14.8} \mev$, $L_4^r(\mu\!=\!770\mev) = +0.38\staterr{09}\syserr{15}\totalerr{18} \times 10^{-3}$ and $L_5^r(\mu\!=\!770\mev) = +0.58\staterr{0.38}\syserr{1.08}\totalerr{1.14} \times 10^{-3}$. For the light quark contribution to the pion scalar radius we obtain $\rsqrpi = 0.515 \staterr{28} \syserr{32} \totalerr{43} \fm^2$, resulting in a value of $\bar{\ell}_4 = 3.99\staterr{15}\syserr{17}\totalerr{23}$ for the pertinent low-energy constant in two-flavor chiral perturbation theory.

 \end{abstract}

 \maketitle

 \section{Introduction} \label{sec:introduction}

In the study of the strong interactions at low energies,
quark-hadron duality dictates that there are essentially two different
approaches that can be taken: either to
describe the strong interactions on the phenomenological level of hadrons, such
as is done in Chiral Perturbation Theory (\chiPT) and its various offshoots, or
to rely on the fundamental quark and gluon degrees of freedom and to determine
hadronic quantities non-perturbatively from the low-energy dynamics of Quantum
Chromodynamics (QCD), such as by means of lattice QCD simulations.

With the advent of much improved algorithms and vastly increased computing
power enabling large-scale simulations at physical quark masses, the working
relationship between lattice QCD and \chiPT has changed: no longer is \chiPT an
essential tool required to extrapolate results from lattice simulations done at
unphysically heavy quark masses to the physical point, but the ability of
lattice simulations to vary the quark masses away from their physical values
allows checking the validity of \chiPT and extracting values for its low-energy
constants (LECs).

Especially well-suited for the latter purpose are quantities that depend only
on one or two LECs, such as the scalar radii
\begin{equation}
 \rsqr{S}{\pi}{f} = \l. - \frac{6}{F_S^{\pi,f}(0)} \frac{dF_S^{\pi,f}(Q^2)}{dQ^2} \r|_{Q^2\rightarrow 0}
 \label{eq:radius}
\end{equation}
parametrizing the slope of the scalar form factors ($Q^2=-q^2=-(p_f-p_i)^2$)
\begin{equation}
 F^{\pi,f}_S(Q^2) = \l\< \pi(\vpf) \r| \mathcal{S}^f \l| \pi(\vpi) \r\> \,,
 \label{eq:scalar_FF}
\end{equation}
where in the theory with $N_f=2$ quark flavors there is only one isoscalar scalar density
\begin{equation}
 \mathcal{S}^l = \bar{u}u + \bar{d}d \,,
 \label{eq:Sl}
\end{equation}
whereas for $N_f=2+1$ quark flavors the basic scalar densities are the singlet and octet ones,
\begin{align}
 \mathcal{S}^0 &= \bar{u}u + \bar{d}d - 2\bar{s}s\label{eq:S8} \,, \\
 \mathcal{S}^8 &= \bar{u}u + \bar{d}d +  \bar{s}s\label{eq:S0} \,.
\end{align}

In $\SU{2}$ \chiPT, the scalar radius depends only on $\bar{\ell}_4$ via
\begin{equation}
 \rsqr{S}{\pi}{l} = \frac{1}{8\pi^2f_{\pi,\phys}^2} \left[-\frac{13}{2}+\bar{\ell}_4+\log\frac{M_{\pi,\mathrm{phys}}^2}{M_\pi^2}\right] \,.
 \label{eq:rsqrl_SU2}
\end{equation}

In $\SU{3}$ \chiPT, the scalar radii are related by \cite{Gasser:1984ux}
\begin{align}
 \rsqr{S}{\pi}{0} &= \rsqr{S}{\pi}{8} + \delta r^2_S, \label{eq:rsqr0} \\
 \rsqr{S}{\pi}{l} &= \rsqr{S}{\pi}{8} + \frac{2}{3}\delta r^2_S, \label{eq:rsqrl}
\end{align}
where
\begin{align}
 \rsqr{S}{\pi}{8} &= \frac{1}{96\pi^2f_0^2}\left[4608\pi^2 L_5^r -60 -2\frac{M_\pi^2}{M_\eta^2}+18\log\frac{M_K^2}{M_\pi^2}+\frac{45M_\pi^2-18M_K^2-27M_\eta^2}{M_K^2-M_\pi^2}\log\frac{M_\pi^2}{\mu^2}\right], \label{eq:rsqr8} \\
 \delta r^2_S &= \frac{12}{f_0^2}\left[12 L_4^r -\frac{3}{64\pi^2}\left(\log\frac{M_K^2}{\mu^2}+1\right)+\frac{1}{288\pi^2}\frac{M_\pi^2}{M_\eta^2} \label{eq:delta_rsqr_S}
 \right].
\end{align}
The octet radius therefore depends solely on $L_5^r$, while the other two radii depend on both $L_5^r$ and $L_4^r$. An accurate determination of the scalar form factors therefore can enable a precise determination of the LECs $\bar{\ell}_4$, and $L_5^r$ and $L_4^r$, respectively. \par

In this paper, we determine the scalar form factors of the pion with high statistics and fully controlled systematics, including the extraction of ground state matrix elements, form factor parametrization and the physical extrapolations. Our form factor data have far better momentum resolution than previous studies due to the use of moving frames, and our determination of the numerically difficult quark-disconnected contribution over the full range of momenta is more than an order of magnitude more accurate than the most precise earlier one. \par 

This paper is organized as follows: Sect.~\ref{sec:ensembles} includes general information on the ensembles and the setup used in this study, whereas technical aspects of the $n$-point function and effective form factor calculation are explained in detail in Sect.~\ref{sec:computational_setup}. The extraction of ground state matrix elements and the parametrization of the form factors to compute radii are discussed in Sect.~\ref{sec:ESA}~and~\ref{sec:FF_parametrization}, respectively. The physical extrapolation of our results using \chiPT{} fit ans\"atze and the determination of LECs is detailed in Sect.~\ref{sec:CCF}. Our final results including a full error budget are obtained from model averages in Sect.~\ref{sec:AIC}. Finally, Sect.~\ref{sec:summary} contains a summary discussion of our findings.
\

 \section{Ensembles} \label{sec:ensembles} 
The lattice calculation of the form factor in Eq.~\ref{eq:scalar_FF} has been carried out on a set of 17 gauge ensembles listed in Table~\ref{tab:ensembles} that are provided by the Coordinated Lattice Simulations (CLS) consortium \cite{Bruno:2014jqa}. These ensembles feature $N_f=2+1$ flavors of non-perturbatively $\mathcal{O}(a)$-improved Wilson fermions \cite{Sheikholeslami:1985ij} and have been generated with the tree-level Symanzik-improved gauge action \cite{Luscher:1984xn}. The majority of ensembles exhibit open boundary conditions in the time direction to mitigate the issue of topological freezing \cite{Luscher:2011kk,Luscher:2012av}, whereas the remaining ensembles feature periodic boundary conditions. In order to suppress exceptional configurations, a twisted-mass regulator \cite{Luscher:2012av} has been introduced in the simulation of the light-quark sector, and the rational approximation has been used for the strange quark \cite{Clark:2006fx}. Therefore, the computation of physical observables requires reweighting, and the corresponding reweighting factors have been determined using exact low-mode deflation as described in Ref.~\cite{Kuberski:2023zky} for all but one of the ensembles in Table~\ref{tab:ensembles}. The only exception is E300, for which reweighting factors are only available from the conventional stochastic method originally introduced in Ref.~\cite{Bruno:2014jqa}. Moreover, we adopt the procedure introduced in Ref.~\cite{Mohler:2020txx} to treat violations of the positivity of the fermion determinant. Such violations have been found to affect a small subset of gauge configurations on some ensembles and can be accounted for in the reweighting procedure by adding a sign to the reweighting factor for a given gauge configuration. In the following, it is implied that the appropriate reweighting has been applied in any gauge average. \par

The majority of ensembles in Table~\ref{tab:ensembles} lies on the same chiral trajectory defined by a constraint on the trace of the quark mass matrix, i.e. $\tr[M] = 2m_l+m_s = \mathrm{const}$, hence denoted by ``$\tr[M]$''. Additional ensembles have been included that belong to a second chiral trajectory defined by a constant value of the strange quark mass $m_s \approx \phys$ denoted by ``$m_s$'' in Table~\ref{tab:ensembles}. Since the difference between $F^{\pi,0}_S(Q^2)$ and $F^{\pi,8}_S(Q^2)$ is entirely mediated by the strange quark contribution, we expect this to improve the handle on the combined light and strange quark mass dependence in chiral extrapolations. The two trajectories (approximately) coincide at the physical point, hence the E250 ensemble is considered part of both trajectories. \par

% This table contains a list of the ensembles used in the charges MK2 project
% STATUS: DONE

\begin{table}[!t]
 \caption{List of gauge ensembles used in this study. Superscripts ``$o$'' and ``$p$'' denote ensembles with open and periodic boundary conditions in time, respectively. The ensembles fall onto two different chiral trajectories labelled by $\tr[M]$ and $m_s$ that are defined in the text. Values for $M_\pi$ and $M_K$ have been computed on the same set of configurations and their errors are dominated by the uncertainty of the scale $\sqrt{t_0}$ in Eq.~(\ref{eq:sqrt_t0phys}) that is used to convert to physical units. The last three columns list the number of gauge configurations $N_\mathrm{cfg}$ that have been analyzed, the total number of measurements for quark-connected three-point functions $N_\mathrm{meas}^\mathrm{3pt}$ and the available number of two-point function measurements $N_\mathrm{meas}^\mathrm{2+1pt}$ used in the computation of the quark-disconnected contributions to the three-point functions.}
 \centering
 \setlength{\tabcolsep}{0.5em}
 \begin{tabular}{cccccccccrrr}
  \hline\hline
  ID$^\mathrm{BC}$ & $\beta$ & $a/\fm$ & $\frac{T}{a}\times\bigl(\frac{L}{a}\bigr)^3$ & traj & $L/\fm$ & $M_\pi L$ & $M_\pi/\mev$ & $M_K/\mev$ & \mc{$N_\mathrm{cfg}$} & \mc{$N_\mathrm{meas}^\mathrm{3pt}$} & \mc{$N_\mathrm{meas}^\mathrm{2+1pt}$} \\
  \hline\hline                                                                                                         % without removing outliers
  C101$^o$ & 3.40 & 0.0855 &  $96 \times 48^3$ & $\tr[M]$ & 4.11 & 4.68 & 225(2) & 476(3) & 1970 & 15760 &  866800 \\  %   2000 & 16000 &  880000
  C102$^o$ &      &        &  $96 \times 48^3$ & $m_s$    & 4.11 & 4.75 & 228(2) & 503(3) & 1467 & 11736 &  645480 \\  %   1500 & 12000 &  660000
  N101$^o$ &      &        & $128 \times 48^3$ & $\tr[M]$ & 4.11 & 5.89 & 283(2) & 466(3) & 1577 & 12616 &  883120 \\  %   1596 & 12768 &  893760
  H102$^o$ &      &        &  $96 \times 32^3$ & $\tr[M]$ & 2.74 & 4.97 & 358(2) & 443(3) & 2029 & 16232 & 1136240 \\  %   2037 & 16296 & 1140720
  \hline
  D450$^p$ & 3.46 & 0.0756 & $128 \times 64^3$ & $\tr[M]$ & 4.84 & 5.36 & 219(1) & 480(3) & 1009 &  8072 &  516608 \\  %   1500 & 12000 &  768000
  D451$^p$ &      &        & $128 \times 64^3$ & $m_s$    & 4.84 & 5.35 & 218(2) & 505(3) & 1482 & 11856 &  758784 \\  %   1028 &  8224 &  526336
  N451$^p$ &      &        & $128 \times 48^3$ & $\tr[M]$ & 3.63 & 5.32 & 289(2) & 465(3) & 1004 &  8032 &  514048 \\  %   1011 &  8088 &  517632
  N452$^p$ &      &        & $128 \times 48^3$ & $\tr[M]$ & 3.63 & 6.50 & 354(2) & 444(3) &  998 &  7984 &  510976 \\  %   1000 &  8000 &  512000
  S400$^o$ &      &        & $128 \times 32^3$ & $\tr[M]$ & 2.42 & 4.37 & 356(2) & 447(3) &  993 &  7944 &  460752 \\  %   1001 &  8008 &  464464
  \hline
  E250$^p$ & 3.55 & 0.0636 & $192 \times 96^3$ & both     & 6.11 & 4.07 & 132(1) & 494(3) &  982 &  7856 &  502784 \\  %   1000 &  8000 &  512000
  D200$^o$ &      &        & $128 \times 64^3$ & $\tr[M]$ & 4.07 & 4.22 & 204(1) & 485(3) & 1984 & 15872 & 1111040 \\  %   2000 & 16000 & 1120000
  D201$^o$ &      &        & $128 \times 64^3$ & $m_s$    & 4.07 & 4.21 & 204(1) & 505(3) & 1062 &  8496 &  594720 \\  %   1078 &  8624 &  603680
  N200$^o$ &      &        & $128 \times 48^3$ & $\tr[M]$ & 3.06 & 4.45 & 287(2) & 467(3) & 1703 & 13624 &  953680 \\  %   1712 & 13696 &  958720
  N203$^o$ &      &        & $128 \times 48^3$ & $\tr[M]$ & 3.06 & 5.40 & 349(2) & 446(3) & 1535 & 12280 &  859600 \\  %   1543 & 12344 &  864080
  \hline
  E300$^o$ & 3.70 & 0.0493 & $192 \times 96^3$ & $\tr[M]$ & 4.74 & 4.22 & 176(1) & 495(3) & 1134 &  9072 &  521640 \\  %   1139 &  9112 &  523940
  J303$^o$ &      &        & $192 \times 64^3$ & $\tr[M]$ & 3.16 & 4.17 & 260(2) & 479(3) & 1068 &  8544 &  598080 \\  %   1073 &  8584 &  600880
  J304$^o$ &      &        & $192 \times 64^3$ & $m_s$    & 3.16 & 4.18 & 261(2) & 526(3) & 1625 & 13000 &  910000 \\  %   1634 & 13072 &  915040
  \hline\hline
 \end{tabular}
 \label{tab:ensembles}
\end{table}

Throughout the analysis dimensionful quantities are expressed in units of the gradient flow scale $t_0$ \cite{Luscher:2010iy} using the values for $t_0^\mathrm{sym}/a^2$ at the symmetrical point as published in Table~III of Ref.~\cite{Bruno:2016plf}. The physical value of $t_0$ enters the analysis only in the definition of the physical point in the light and strange quark mass and in the conversion of final results to physical units. For the purpose of setting the scale we use the FLAG world average estimate for $N_f=2+1$ dynamical quark flavors \cite{FlavourLatticeAveragingGroupFLAG:2021npn}
\begin{equation}
 \sqrt{t_0^\phys}=0.14464(87)\fm \,,
 \label{eq:sqrt_t0phys}
\end{equation}
Note that the values for the lattice spacings $a$ in Table~\ref{tab:ensembles} have been computed using the values for $\sqrt{t_0^\phys}$ and $t_0^\mathrm{sym}/a^2$. They have only been used for the purpose of converting $\Mpi$ and $\MK$ in Table~\ref{tab:ensembles} to physical units and do not explicitly enter the actual analysis. \par

 \section{Computational setup} \label{sec:computational_setup}

Evaluating the matrix elements in Eq.~(\ref{eq:scalar_FF}) in lattice QCD requires the calculation of two- and three-point functions
\begin{align}
 \Ctwopt{PP}{\vp, \vxi, t_f-t_i} &= \sum_{\vxf} e^{i \vp\cdot(\vxf - \vxi)} \< P(\vxf-\vxi, t_f-t_i) P^\dag(\mathbf{0}, 0) \>_F \,, \label{eq:2pt} \\
 \Cthreept{P}{S}{f}{\vpf, \vq, \vxi, t_f-t_{i}, t_{op}-t_i} &= \sum_{\vxf,\vxop} e^{i \vpf \cdot (\vxf-\vxi)} e^{i\vq \cdot (\vxop-\vxi)} \< P(\vxf-\vxi, t_f-t_i) \mathcal{S}^f(\vxop-\vxi, t_{op}-t_{i}) P^\dag(\mathbf{0}, 0) \>_F \,, \label{eq:3pt}
\end{align}
where $P(\vx, t) = \frac{1}{\sqrt{2}} \l[\bar{u}\g{5}u + \bar{d}\g{5}d\r](\vx,t)$ is the usual pseudoscalar interpolating field for the pion, and $\< \cdot \>_F$ denotes fermionic expectation values. Coordinates of initial and final states are labeled by indices ``$i$'' and ``$f$'', respectively, and the location of the operator insertion is indicated by ``$op$''. In the above expressions a shift by the source position $x_i$ has already been carried out assuming translational invariance. \par

\begin{figure}[t]
 \includegraphics[width=.9\textwidth]{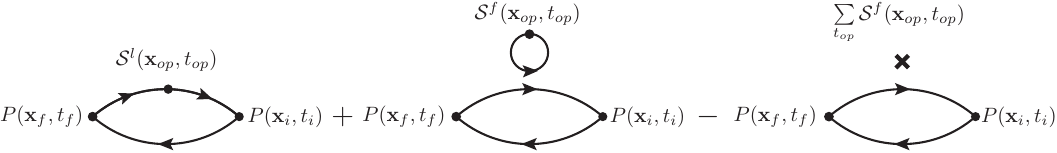}
 \caption{Quark-connected and disconnected diagrams contributing to the three-point function in Eq.~(\ref{eq:3pt}). The quark-connected diagram receives only light-quark contributions, while the quark-disconnected contribution depends on the choice of the flavor structure for the operator insertion in Eqs.~(\ref{eq:S8})--(\ref{eq:Sl}). The last diagram accounts for the subtraction of the vacuum expectation value that occurs only at vanishing momentum transfer.}
 \label{fig:3pt}
\end{figure}

The Wick contractions for the three-point function in Eq.~(\ref{eq:3pt}) give rise to quark-connected and quark-disconnected contributions that are diagrammatically depicted in Fig.~\ref{fig:3pt}. In the presence of pion initial and final states the quark-connected diagram contributes only for operator insertions involving light quarks. The quark-disconnected piece, on the other hand, contributes for any quark flavor. Moreover, the scalar current insertion in Eqs.~(\ref{eq:S0})--(\ref{eq:Sl}) has vacuum quantum numbers, hence there is an additional contribution at zero momentum transfer due to a vacuum expectation value (vev) that requires subtraction to obtain physical results, which will be discussed in more detail in Subsection~\ref{subsec:vev_subtraction}. \par

\subsection{Quark-connected contribution} \label{subsec:connected_diagrams}
The quark-connected piece of the three-point function in Fig.~\ref{fig:3pt} is computed to high statistical precision using only a few point sources per gauge field. We employ sequential inversion through the sink \cite{LHPC:2002xzk} and the truncated solver method (TSM) \cite{Bali:2009hu,Blum:2012uh,Shintani:2014vja}. The computations are carried out for a fixed set of source-sink separations $\tsep=t_f-t_i$ (assuming $t_f>t_i)$ in lattice units at each value of $\beta$ as listed in Table~\ref{tab:beta_dependent_parameters}. They are chosen such that always they cover a similar range of physical values, i.e. $\tsep\in[\tseplo,\tsephi]$ with $\tseplo\lesssim1\fm$ and $\tsephi\gtrsim3\fm$. The final state is either produced at rest or with one unit of momentum, i.e. $\vpf=(+1,0,0)\cdot 2\pi/L$, which greatly improves the momentum resolution. Since the injection of momenta at the sink requires additional sequential inversions for every single value of $\vpf$, we restrict ourselves to a single non-vanishing momentum in case of the quark-connected three-point function. \par

% STATUS: XCHECK

\begin{table}[!t]
 \caption{Lower and upper bounds $\tseplo$ and $\tsephi$ as well as the stride between to consecutive values of $\tsep$ for which three-point function data has been computed.}
 \centering
 \setlength{\tabcolsep}{0.5em}
 \begin{tabular}{cccc}
  \hline\hline
  $\beta$ & $\tseplo/a$ & $\tsephi/a$ & $\Delta\tsep/a$ \\
  \hline\hline
   3.40 & 11 & 35 & 4 \\
   3.46 & 13 & 41 & 4 \\
   3.55 & 15 & 51 & 6 \\
   3.70 & 19 & 67 & 8 \\
  \hline\hline
 \end{tabular}
 \label{tab:beta_dependent_parameters}
\end{table}

The source setup on individual ensembles depends on the choice of boundary conditions in time: For periodic boundary conditions (pBC) the quark-connected diagrams are evaluated on $N_s^c=8$ point sources that are randomly selected on every gauge field. The point-to-all propagators are re-used for each value of $\tsep$ and in the computation of two-point functions. For ensembles with open boundary conditions (oBC) the source setup is chosen symmetric in $t$ around the center of the lattice with two opposing source timeslices $t_i=(T-1\pm\tsep)/2$ at each given value of $\tsep$. This choice restricts the possible source-sink separations to odd values of $\tsep/a$, which is reflected by the values in Table~\ref{tab:beta_dependent_parameters}. We use four point sources on both timeslices, averaging over forward and backward diagrams in insertion time $\tins=t_{op}-t_i$, such that the number of sources per configuration and source-sink separation remains the same as for pBC ensembles, i.e. $N_s^c=8$. However, the number of point-to-all propagators now scales with the number of source-sink separations. Again, these are re-used for the computation of two-point functions. The total numbers of measurements $N_\mathrm{meas}^\mathrm{3pt}$ for the quark-connected piece of the three-point function on every ensemble are listed in Table~\ref{tab:ensembles}. \par

The low-precision solves for two- and three-point functions are averaged over source positions $x_i$. For ensembles with pBC this can be done without any restrictions regarding the source time $t_i$, and we correct for the associated bias on every gauge field using a single high-precision solve on a randomly chosen source position $x_i^b$, $b \in \{1, \ldots, N_s^c \}$, i.e.
\begin{align}
 \Ctwopt{PP}{\vp, t} &= \CtwoptHP{PP}{\vp, \vxi^b, t_f^b-t_i^b} - \CtwoptLP{PP}{\vp, \vxi^b, t_f^b-t_i^b} + \frac{1}{N_s^c} \sum\limits_{j=1}^{N_s^c} \CtwoptLP{PP}{\vp, \vxi^j, t_f^j-t_i^j} \,, \label{eq:2pt_x_i_avg_pBC} \\
 \Cthreept{P}{S}{f}{\vpf, \vq, \vpi, \tsep, \tins} &= \CthreeptHP{P}{S}{f}{\vpf, \vq, \vxi, t_f^b-t_i^b, t_{op}^b-t_i^b} - \CthreeptLP{P}{S}{f}{\vpf, \vq, \vxi, t_f^b-t_i^b, t_{op}^b-t_i^b} \notag \\
  & \quad + \frac{1}{N_s^c} \sum\limits_{j=1}^{N_s^c} \CthreeptLP{P}{S}{f}{\vpf, \vq, \vxi, t_f^j-t_i^j, t_{op}^j-t_i^j} \,, \label{eq:3pt_x_i_avg_pBC}
\end{align}
where we have introduced the usual time indexing $t=t_f^j-t_i^j$ for the two-point function and made the $\vpi$-dependence of the three-point function explicit in its argument list. Superscripts ``$\mathrm{HP}$'' and ``$\mathrm{LP}$'' refer to high- and low-precision solves, respectively. On the other hand, for ensembles with oBC the source average is performed at any given value of $\tsep$, averaging only over the forward and backward contributions in $\tins$ obtained from the two contributing source timeslices with $t_i=(T-1\pm\tsep)/2$. The bias correction is carried out on a randomly selected (spatial) source for each of the two values of $t_i$ on every individual gauge configuration.

For the extraction of effective form factors we average over equivalent momenta before forming ratios. To this end, we define equivalence classes $\mfp=\l\{\tilde{\vp}\in\mathbb{P}^3 \mid \tilde{\vp}^2=\vp^2\r\}$, where $\mathbb{P}^3=\l\{\vp=\frac{2\pi}{L}\mathbf{n} \mid \mathbf{n}\in\mathbb{Z}^3\r\}$ denotes the set of lattice three-momenta. Carrying out the corresponding momentum average for the two-point function in Eq.~\ref{eq:2pt} yields
\begin{equation}
  \Ctwopt{PP}{\mfp, t} = \frac{1}{N_\mfp} \sum\limits_{\vp\in\mfp} \Ctwopt{PP}{\vp, t} \,,
  \label{eq:2pt_averaged_pBC}
\end{equation}
where $N_\mfp$ denotes the number of distinct lattice momenta $\vp$ in the class $\mfp$, and we have defined $t=t_f-t_i$. Similarly, for three-point functions we introduce classes of momentum triplets $\l(\mfpf,\mfq,\mfp_i\r) = \{\l(\tilde{\vp}_f,\tilde{\vq},\tilde{\vp}_i\r) \in \mathbb{P}^3 \times \mathbb{P}^3 \times \mathbb{P}^3 \mid (\tilde{\vp}_f^2,\tilde{\vq}^2,\tilde{\vp}_i^2) = (\vpf^2, \vq^2, \vpi^2)\}$. Since they are unambiguously defined by triplets of squared lattice momenta it is convenient to introduce corresponding labels $\mathcal{P}=\{n(\vpf^2), n(\vq^2), n(\vpi^2)\}$ with $n(\vec{p}^2)=\vec{p}^2\cdot (L/2\pi)^2$. Again, we average the three-point functions over equivalent momenta and source positions $x_i$
\begin{equation}
 \Cthreept{P}{S}{f}{\mfpf, \mfq, \mfpi, \tsep, \tins} = \frac{1}{N_\mathcal{P}} \sum\limits_{(\vpf,\vq,\vpi)\in(\mfpf,\mfq,\mfpi)} \Cthreept{P}{S}{f}{\vpf, \vq, \vpi, \tsep, \tins} \,,
 \label{eq:3pt_averaged}
\end{equation}
where $N_\mathcal{P}$ denotes the multiplicity of momenta in any given class $(\mfpf,\mfq,\mfpi)$. \par

The computation of the effective form factor requires the cancellation of unknown overlap factors, which can be achieved in the standard way by forming a ratio of two- and three-point functions
\begin{equation}
 R^f(\mfpf, \mfq, \mfpi, \tsep, \tins) = \frac{\<\Cthreept{P}{S}{f}{\mfpf,\mfq,\mfpi, \tsep, \tins}\>}{\<\Ctwopt{PP}{\mfpf, \tsep}\>} \sqrt{\frac{\<\Ctwopt{PP}{\mfpi, \tsep-\tins}\> \<\Ctwopt{PP}{\mfpf^2, \tins}\> \<\Ctwopt{PP}{\mfpf^2, \tsep}\>}{\<\Ctwopt{PP}{\mfpf^2, \tsep-\tins}\> \<\Ctwopt{PP}{\mfpi^2, \tins}\> \<\Ctwopt{PP}{\mfpi^2, \tsep}\>}} \,,
 \label{eq:ratio}
\end{equation}
where $\<.\>$ indicates gauge averages. Note that for a scalar operator insertion the form factor decomposition is trivial, i.e. $R^f(\mfpf, \mfq, \mfpi, \tsep, \tins)$ up to excited state contamination directly yields the unrenormalized matrix elements without additional kinematic structures. For the quark-connected contribution to $\Cthreept{P}{S}{l}{\mfpf,\mfq,\mfpi, \tsep, \tins}$ a favorable signal-to-noise behavior for $R^l(\mfpf, \mfq, \mfpi, \tsep, \tins)$ is obtained by evaluating two- and three-point functions on the exact same set of sources, as opposed to using full available statistics for the two-point functions. In particular, for ensembles with oBC we neither average over source positions $t_i$ contributing to different values of $\tsep$ in the quark-connected part of $\Cthreept{P}{S}{l}{\mfpf,\mfq,\mfpi, \tsep, \tins}$, nor do we include additional two-point function measurements required for the computation of (2+1)-point quark-disconnected diagrams that will be discussed below. To further optimize correlations for the evaluation of Eq.~(\ref{eq:ratio}), the two-point functions are not averaged over forward and backward time direction, but we rather choose the time direction that coincides with the $\tins$-indexing of $\Cthreept{P}{S}{f}{\mfpf,\mfq,\mfpi, \tsep, \tins}$, i.e. forward direction on ensembles with pBC and forward (backward) direction on ensembles with oBC for $t_i<T/2$ ($t_i>T/2$), as explained before. \par

Statistical errors for any observable are always computed and propagated using the non-parametric bootstrap method with $N_B=1000$ bootstrap samples and binning to account for autocorrelation. Similar to what has been observed previously for isoscalar nucleon structure observables in Refs.~\cite{Agadjanov:2023efe,Djukanovic:2023beb,Djukanovic:2023jag}, a few measurement on individual point sources show up as extreme outliers in the corresponding distribution across gauge configurations. While this affects only very small subset of gauge fields, these measurements would spoil the expected scaling behavior of the statistical error with the total number of measurements as well as the expected signal-to-noise ratio as a function of $\tsep$, $\tins$. Moreover, including them causes a generally unreasonable increase of statistical errors. Therefore, we have implemented a similar procedure as the one discussed in the supplemental material of Ref.~\cite{Agadjanov:2023efe} to flag these measurements and remove them from the computation of gauge averages. For this purpose, we generate single-elimination jackknife samples for the effective form factors on the full set of data and scan them for results that are further away than $\sim6\sigma$ from the center of the sample distribution. Any gauge configurations identified by this procedure is then removed from the actual (bootstrap-based) analysis. This elimination is already reflected by the number of gauge configurations $N_\mathrm{conf}$ listed in Table~\ref{tab:ensembles}. \par

\subsection{Quark-disconnected contribution} \label{subsec:disconnected_diagrams}
The numerical evaluation of the quark-disconnected contribution to Eq.~(\ref{eq:3pt}) requires the computation of two-point functions $\Ctwopt{PP}{\vpf, \vxi, t}$ and scalar quark loops
\begin{equation}
 L_{\mathcal{S}^f}(\vq, t) = - \sum\limits_{\vx} e^{i \vq \cdot \vx} \trace{D_f^{-1}(x,x)} \,,
\end{equation}
where $D_f^{-1}(x,x)$ denotes the self-contracted all-to-all propagator for a quark flavor $f=l,s$. The computation of these loops has been carried out on all ensembles in the scheme that we have introduced in Ref.~\cite{Ce:2022eix} based on the method in Ref.~\cite{Giusti:2019kff}, combining the one-end trick (OET) \cite{McNeile:2006bz}, the generalized hopping parameter expansion (gHPE) \cite{Gulpers:2013uca} and hierarchical probing (HP) \cite{Stathopoulos:2013aci}. In this scheme quark loops are computed for any local bilinear or one-link displaced operator without incurring additional inversion cost for individual operators, facilitating their (re-)use in various other projects; e.g. in nucleon structure calculations in Refs.~\cite{Agadjanov:2023efe,Djukanovic:2023beb,Djukanovic:2023jag} as well as the hadronic vacuum polarization and hadronic light-by-light contributions to $g-2$ in Refs.~\cite{Chao:2021tvp,Ce:2022kxy,Djukanovic:2024cmq}. \par

The measurements for the pion two-point function that are obtained from the forward propagators of the quark-connected three-point functions are generally insufficient to achieve a statistically precise signal for the quark-disconnected contribution in Fig.~\ref{fig:3pt}. Therefore, we have carried out a dedicated effort to increase statistics for the two-point functions. Again, the numerical setup for these additional measurements as well as evaluation of the (2+1)-point contributions depends on the choice of the boundary conditions in time. For pBC ensembles additional source positions can be randomly distributed on each configuration such that the number of measurement per gauge configuration is increased to $N_s^d=512$ on every ensemble. Unlike for the quark-connected piece, forward- and backward-averaging in $\tins$ does not incur additional cost for inversions. Therefore, we increase statistics for any source position $x_i$ of the two-point function by averaging over the two contributions
\begin{align}
 \Cthreeptdisc{P}{S}{f}{\vpf, \vq, \vpi, \vxi, \tsep, \tins}     &= \Ctwopt{PP}{\vpf, \vxi, \tsep}   \tilde{L}_{\mathcal{S}^f}(\vq, \tins)   e^{-i\vq\vxi} \,, \label{eq:3pt_disc_fw} \\
 \Cthreeptdisc{P}{S}{f}{\vpf, \vq, \vpi, \vxi, T-\tsep, T-\tins} &= \Ctwopt{PP}{\vpf, \vxi, T-\tsep} \tilde{L}_{\mathcal{S}^f}(\vq, T-\tins) e^{-i\vq\vxi} \,, \label{eq:3pt_disc_bw}
\end{align}
where the indexing of the loop has been assumed to be implicitly shifted by $t_i$, i.e. $\tilde{L}_{\mathcal{S}^f}(\vq,t)=L_{\mathcal{S}^f}(\vq,t+t_i)$. The implementation of the TSM for the two-point function in Eq.~(\ref{eq:2pt_x_i_avg_pBC}) is straightforwardly extended to the (2+1)-point case. Since the final state momentum $\vpf$ in Eqs.~(\ref{eq:3pt_disc_fw})~and~(\ref{eq:3pt_disc_bw}) is determined by the two-point function contribution, all signed permutations contributing to one unit of final state momentum can be computed without additional inversions, increasing the statistical precision for the resulting signal. The computation of the ratio in Eq.~(\ref{eq:ratio}) for any flavor structure $f$ then proceeds in the same way as for the quark-connected piece by computing gauge averages and averaging two- and three-point functions over equivalent momenta as discussed in subsection~\ref{subsec:connected_diagrams}. In order to optimize correlations in $R^f(\mfpf, \mfq, \mfpi, \tsep, \tins)$ we evaluate the (2+1)-point and two-point functions on exactly the same (extended) set of 512 source positions. Moreover, we average also the two-point functions in the ratio themselves over forward and backward contributions in $t$. \par

% STATUS: DONE

\begin{table}[!t]
 \caption{Values of $\tex$ defining the exclusion regions next to the boundaries in time for ensembles and the number of available two-point function measurements for the construction of (2+1)-point quark-disconnected diagrams $N_s^d$ on these ensembles.}
 \centering
 \setlength{\tabcolsep}{0.5em}
 \begin{tabular}{ccc}
  \hline\hline
  ID   & $\tex/a$ & $N_s^d$ \\
  \hline\hline
  C101 & 26 & 440 \\ %% 2.25fm
  C102 & 26 & 440 \\ %% 2.25fm
  N101 & 23 & 548 \\ %% 2.00fm
  H102 & 20 & 548 \\ %% 1.75fm
  \hline
  S400 & 23 & 464 \\ %% 1.75fm
  \hline
  D200 & 35 & 548 \\ %% 2.25fm
  D201 & 35 & 548 \\ %% 2.25fm
  N200 & 31 & 548 \\ %% 2.00fm
  N203 & 27 & 548 \\ %% 1.75fm
  \hline
  E300 & 50 & 448 \\ %% 2.50fm
  J303 & 40 & 548 \\ %% 2.00fm
  J304 & 40 & 548 \\ %% 2.00fm
  \hline\hline
 \end{tabular}
 \label{tab:oBC_related_parameters}
\end{table}

\begin{figure}[t]
 \centering
 \includegraphics[totalheight=0.3\textheight]{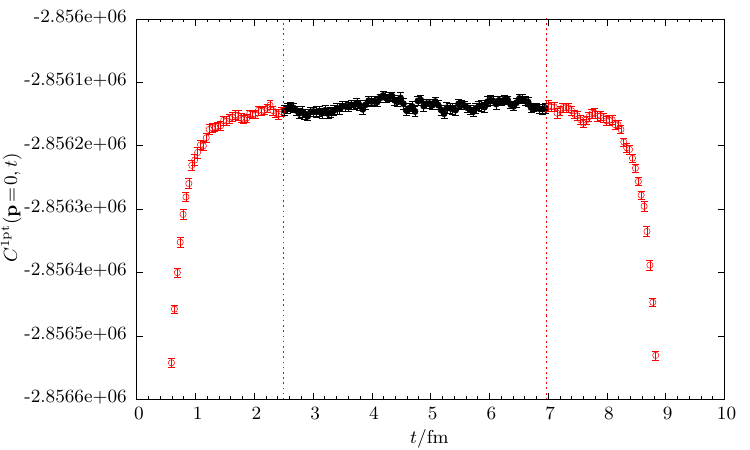}
 \caption{Time-dependence of the scalar one-point function with non-vanishing vacuum expectation value on the E300 ensemble with open boundary conditions. The vertical dashed lines mark the exclusion zone next to the boundaries (see text), and the red data are discarded in the computation of quark-disconnected (2+1)-point functions to avoid excited-state contamination due to boundary effects.}
 \label{fig:1pt_oBC}
\end{figure}

For ensembles with oBC, the evaluation of the quark-disconnected diagrams and the placement of additional sources for two-point functions become more intricate as boundary effects must be avoided in the construction of $n$-point functions. First of all, they result in excited state contamination in e.g. the scalar one-point function at zero-momentum $L_{\mathcal{S}^f}(\vec{0}, t)$ as shown in Fig.~\ref{fig:1pt_oBC} for the E300 ensemble. The regions where the signal for the vev $L_{\mathcal{S}^f}(\vec{0}, t)$ exhibits a non-constant $t$-dependence must be excluded for the computation of the quark-disconnected diagrams, i.e. we restrict $L_{\mathcal{S}^f}(\vp, t)$ to $t_i\in\l[\tex, T-\tex\r]$. The values of $\tex$ generally depend on $M_\pi$ and have been listed in Table~\ref{tab:oBC_related_parameters}. Similarly, we restrict the timeslices for two-point function sources to $t_i\in\l[\tex, T-\tex\r]$. Statistics for the two point function are increased by incrementally adding pairs of timeslices starting at $\tex$ and $T-\tex$ towards the center of the lattice until the target statistics of $N_s^d\geq512$ per configuration is reached or no more timeslices are available, cf. Table~\ref{tab:oBC_related_parameters}. For ensembles at $\beta=3.7$ (and S400) we use only every second timeslice in this procedure, whereas for all other ensembles every value of $t_i$ is allowed. Again, the bias correction for the TSM has been performed on a single, randomly selected (spatial) source every value of $t_i$. Besides, we carry out additional measurements on the timeslices used in the computation of the quark-connected contribution such that the number of two-point function measurements per timeslice is the same for every value of $t_i$. \par

For the evaluation of Eqs.~(\ref{eq:3pt_disc_fw})~and~(\ref{eq:3pt_disc_bw}) not only the source timeslice $t_i$ of $\Ctwopt{PP}{\vpf, \vxi, t_f-t_i}$ but also $t_f$ must remain within $\l[\tex,T-\tex\r]$. This implies an additional, severe constraint on the effective statistics as $t_i$ is restricted to two strips, i.e. $t_i\in\l[\tex,T-\tex-\tsep\r]$ and $t_i\in\l[\tex+\tsep,T-\tex\r]$ which linearly shrink for increasing values of $\tsep$. Therefore, the values of $N_s^d$ as listed in Table~\ref{tab:oBC_related_parameters} are merely an upper bound on the actual statistics. Anyhow, we still average over forward and backward contributions in $\tins$ whenever the choice of $\tsep$ and $t_i$ allows for it, but the gain of statistics from this is very limited compared to the case of pBC in time. \par

\subsection{Vacuum expectation value subtraction at \texorpdfstring{$\vec{q}=0$}{q=0}} \label{subsec:vev_subtraction}
\begin{figure}[t]
 \centering
 \includegraphics[totalheight=0.226\textheight]{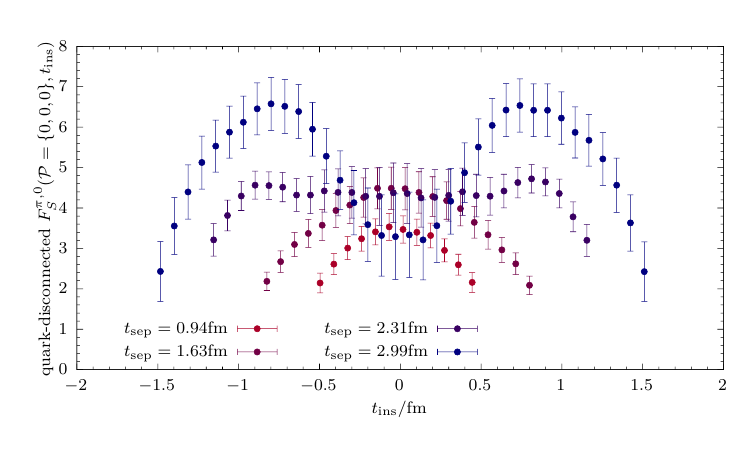}
 \includegraphics[totalheight=0.226\textheight]{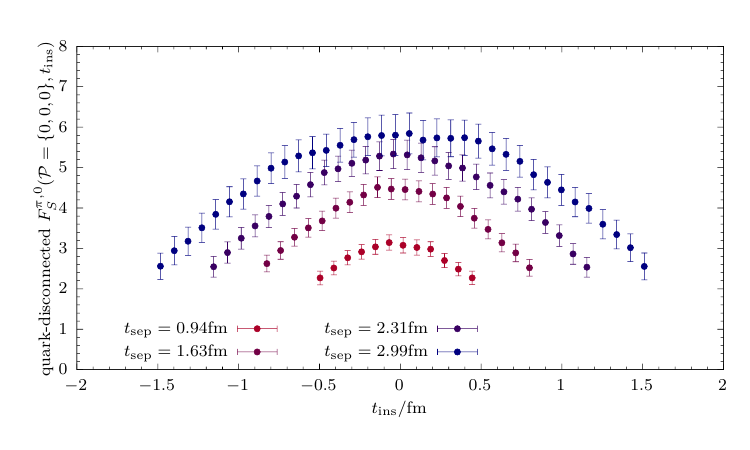} \\
 \includegraphics[totalheight=0.226\textheight]{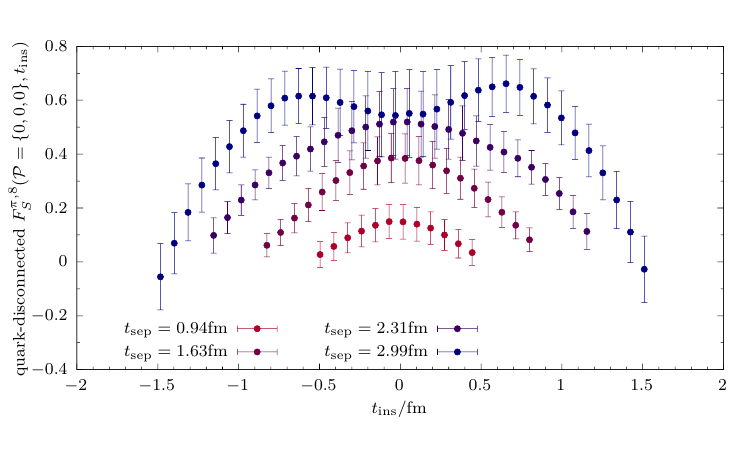}
 \includegraphics[totalheight=0.226\textheight]{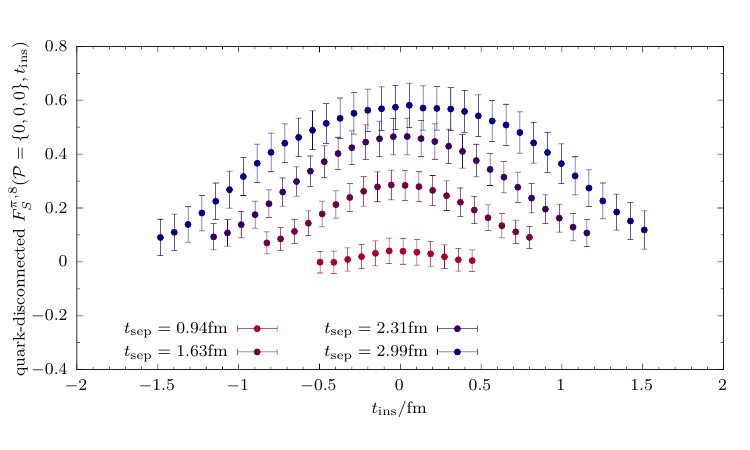}  
 \caption{Comparison of the quark-disconnected contribution to the (unrenormalized) effective form factor at vanishing momentum transfer using the two different methods for the subtraction of the vacuum expectation value on the C101 ensemble with open boundary conditions. Left column: ``naive'' subtraction as defined in Eq.~(\ref{eq:naive_vev_subtraction_oBC}); right column: improved, timeslice-wise method as defined in Eq.~(\ref{eq:improved_vev_subtraction_oBC}). Results are shown for the flavor singlet (top row) and octet (bottom row) combinations. Data at every second available value of $\tsep$ have been included and displaced horizontally for better legibility.}
 \label{fig:vev_subtraction_oBC}
\end{figure}

At vanishing momentum transfer the scalar quark loops feature a non-vanishing vev $\< L_{\mathcal{S}^f}(\vq=\vec{0}, t) \> \neq 0$ that requires subtraction to obtain physical results. Assuming pBC in time, this can achieved in the standard way, i.e. by subtracting
\begin{equation}
 \< \Cvev{P}{S}{f}{\vpf, \vec{q}=\vec{0}, \tsep} \>_\mathrm{pBC} = \< \Ctwopt{PP}{\vpf, \tsep} \> \cdot \frac{1}{T}\sum_{t=0}^{T-1}\< L_{\mathcal{S}^f}(\vec{0}, t) \> \,,
 \label{eq:vev_subtraction_pBC}
\end{equation}
from the r.h.s of Eqs.~(\ref{eq:3pt_disc_fw})~and~(\ref{eq:3pt_disc_bw}) for $\vec{q}=\vec{0}$, $\vpf=\vpi$ after taking the appropriate source and gauge averages. In principle, a similar expression could be used for oBC in time
\begin{equation}
 \< \Cvev{P}{S}{f}{\vpf, \vec{q}=\vec{0}, \tsep} \>_\mathrm{oBC} = \< \Ctwopt{PP}{\vpf, \tsep} \> \cdot \frac{1}{T-2\tex}\sum_{t=\tex}^{T-\tex-1}\< L_{\mathcal{S}^f}(\vec{0}, t) \>
 \label{eq:naive_vev_subtraction_oBC}
\end{equation}
excluding the regions affected by boundary effects in the $t$-average of the loop. However, we observe that this naive ansatz leads to unacceptably large fluctuations in the effective form factor signal on several ensembles. An example for this is shown in Fig.~\ref{fig:vev_subtraction_oBC} for the flavor singlet and octet quark-disconnected contributions to the effective form factor on the C101 ensemble. This effect is caused by insufficient sampling of the quark loop in the evaluation of the quark-disconnected contribution due to the aforementioned restrictions on the placement of point sources for the two-point functions. As a result, local fluctuations in the signal of the loop in $t$ are not adequately averaged out. On the other hand, subtracting the ``naive'' estimator for the vev in Eq.~(\ref{eq:naive_vev_subtraction_oBC}) does not compensate such fluctuations but only amounts to a constant shift of the signal for the resulting three-point function. Since the vev of the loop is several orders of magnitude larger than the signal of the final three-point function these fluctuations may easily dominate the signal, hindering further analysis. The issue becomes increasingly more severe at larger values of $\tsep$ as the available source timeslices are further reduced. For example, at the largest value of $\tsep=35a$ on the C101 ensemble the size of the two strips next to the regions excluded for source placement in $t$ shrink to $\lesssim 0.4\fm$ amounting to only a handful of timeslices for the two-point functions. \par

A remedy for these fluctuations is found by implementing a modified estimator for the effective form factor at $\vq^2=0$. Instead of computing global gauge averages of Eqs.~(\ref{eq:3pt_disc_fw})~and~(\ref{eq:3pt_disc_bw}) including any source position $x_i$, we restrict them to individual source timeslices $t_i$. The corresponding expressions $\<\Cthreeptdisc{P}{S}{f}{\vpf, \vq, \vpi, t_i, \tsep, \tins}\>$ and $\<\Cthreeptdisc{P}{S}{f}{\vpf, \vq, \vpi, t_i, T-\tsep, T-\tins}\>$ are averaged only over spatial source positions $\vxi$ on the same source timeslice $t_i$. Consequently, we introduce a modified estimator for the vev subtraction
\begin{equation}
 \< \Cvev{P}{S}{f}{\vec{q}=\vec{0}, t_i, \tsep, \tins} \>_\mathrm{oBC} = \< \Ctwopt{PP}{\vpf, t_i, \tsep} \> \< \tilde{L}_{\mathcal{S}^f}(\vec{0}, \tins) \> \,,
 \label{eq:improved_vev_subtraction_oBC}
\end{equation}
i.e. we perform the subtraction for every value of $t_i$ separately and without an average in the operator insertion time. This correlated subtraction leads to a cancellation of residual fluctuations in the quark-loop time dependence. The final average over source timeslices $t_i$ is only carried out after building the ratio. Specifically, we replace the quark-disconnected contribution to the ratio in Eq.~(\ref{eq:ratio}) by
\begin{equation}
 R^f_\mathrm{disc}(\mfpf, \mfq=\vec{0}, \mfpi=\mfpf, \tsep, \tins) = \frac{1}{N_{t_i}(\tsep)} \sum_{t_i} \frac{\<\Cthreept{P}{S}{f}{\mfpf,\mfq=\vec{0},\mfpi=\mfpf, t_i, \tsep, \tins}\>_\mathrm{disc}}{\<\Ctwopt{PP}{\mfpf, t_i, \tsep}\>} \,,
 \label{eq:ratio_zero_mom_oBC}
\end{equation}
where $N_{t_i}(\tsep)$ denotes the number of contributing source timeslices at any given value of $\tsep$, and gauge averages of the two-point function $\<\Ctwopt{PP}{\mfpf, t_i, \tsep}\>$ have been computed per timeslice as well. As can be seen comparing the right panels of Fig.~\ref{fig:vev_subtraction_oBC} to the ones on the left, this procedure entirely removes the fluctuations that otherwise distort the signal for the quark disconnected contribution to the form factor. Moreover, it leads to a significant reduction of the point error by a factor up to two, exploiting statistical correlations timeslice-by-timeslice. \par

 \section{Extraction of ground state matrix elements} \label{sec:ESA}
\begin{figure}[t]
 \centering
 \includegraphics[totalheight=0.226\textheight]{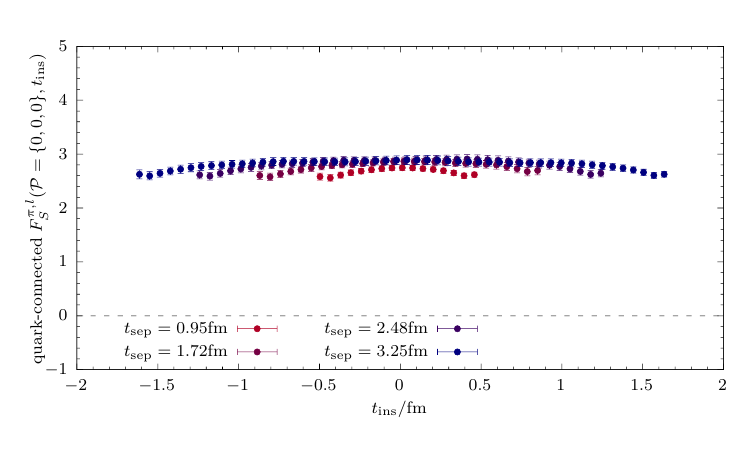}
 \includegraphics[totalheight=0.226\textheight]{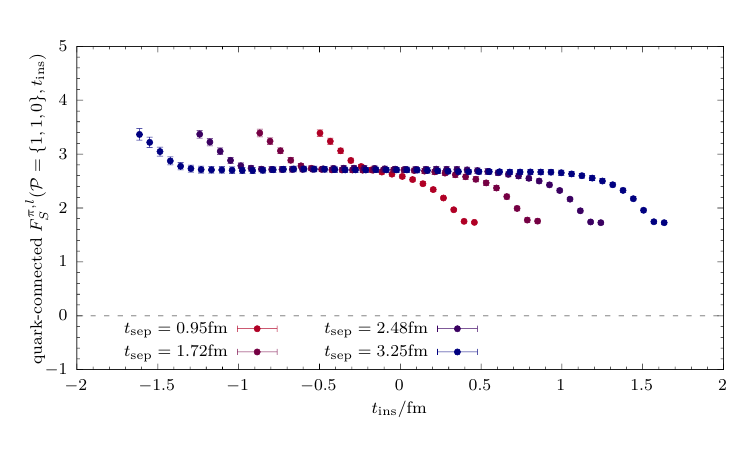} \\
 \includegraphics[totalheight=0.226\textheight]{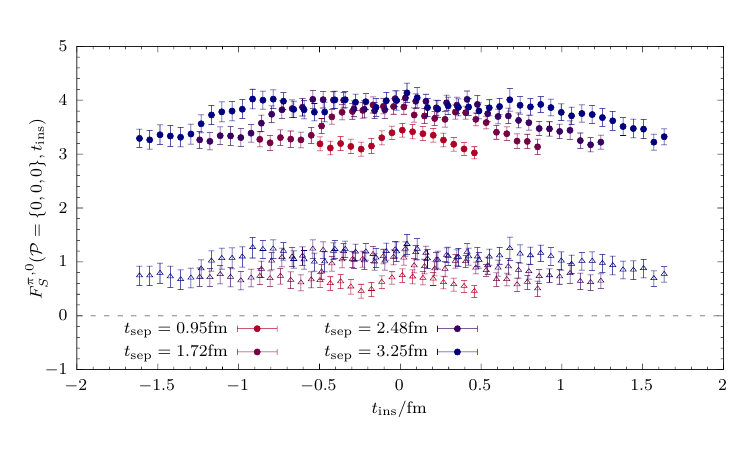}
 \includegraphics[totalheight=0.226\textheight]{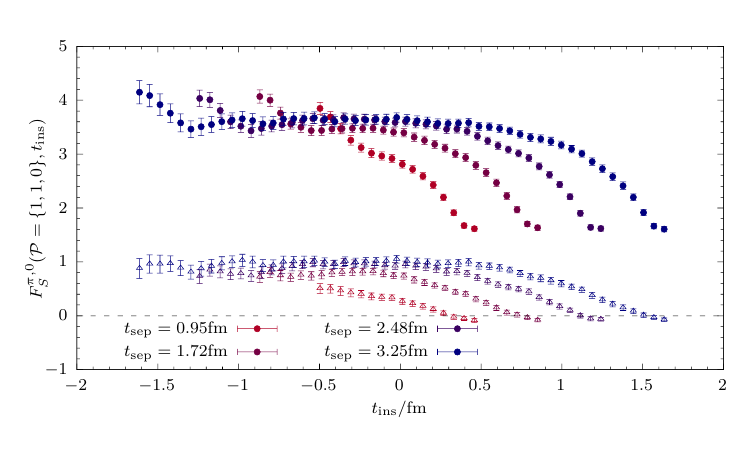} \\
 \includegraphics[totalheight=0.226\textheight]{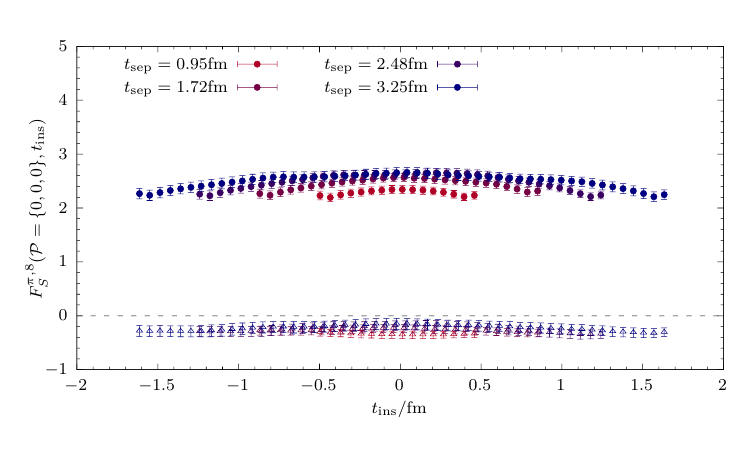}
 \includegraphics[totalheight=0.226\textheight]{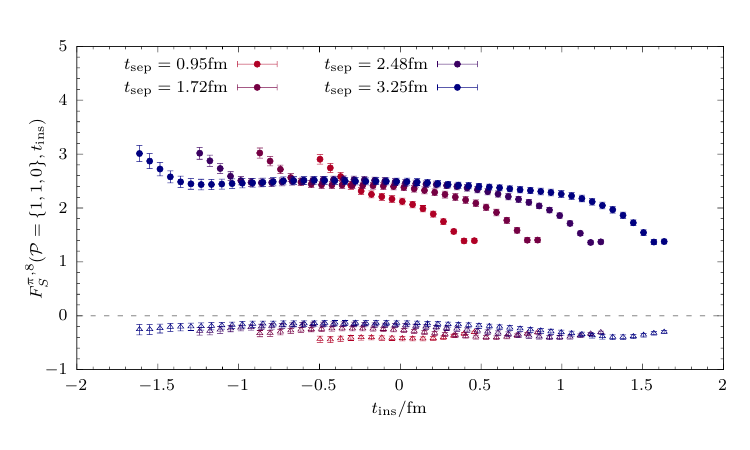}
 \caption{Effective form factors on the E250 ensemble for every second available value of $\tsep$. Results are shown (from top to bottom) for the quark-connected, the flavor singlet and octet combinations. For the flavor singlet and octet combinations the full signal (filled symbols) and the quark-disconnected contribution (open symbols) have been included in the same panel. $\mathcal{P}=\l\{\vpf^2,\vq^2,\vpi^2\r\}$ indicates the momentum configuration, i.e. vanishing momentum transfer and sink momentum (left column), and one unit of momentum transfer at non-vanishing sink momentum (right column). Data at different values of $\tsep$ have been displaced horizontally for better legibility.}
 \label{fig:eff_FF}
\end{figure}

The extraction of ground-state matrix elements for the pion scalar form factor in Eq.~(\ref{eq:scalar_FF}) from lattice data requires control over excited state contamination. For this purpose we employ the summation method~\cite{Maiani:1987by,Gusken:1989ad,Bulava:2011yz,Capitani:2012gj}
\begin{equation}
 \sum\limits_{\tins=\tau}^{\tsep-\tau} R^f(\mfpf, \mfq, \mfpi, \tsep, \tins) = \mathrm{const} + \l\< \pi(\mfpf) \r| \mathcal{S}^f \l| \pi(\mfpi) \r\> (\tsep-\tau) + \mathcal{O}\l(e^{-\Delta\tsep}\r) \,, 
 \label{eq:summation_method}
\end{equation}
where $\Delta$ denotes the energy gap between ground and first excited state. The matrix element is extracted from a correlated linear fit in $\tsep\in\l[\tsepmin,\tsepmax\r]$, where we choose $1.0\fm\lesssim\tsepmin\lesssim1.5\fm$ and $2.5\fm\lesssim\tsepmax\lesssim3.0\fm$, using three values for $\tsepmin$ and $\tsepmax$ each. The remaining analysis is carried out for all resulting $\l(\tsepmin,\tsepmax\r)$--pairs to account for systematic effects due to possible residual excited state contamination at smaller values of $\tsep$, as well as potential issues due the decreasing statistical precision of the quark-disconnected piece at large values of $\tsep$ in case of ensembles with oBC in time. Furthermore, we always choose $\tex=\tsepmin/2$, i.e. the most restrictive choice possible, which we find to yield the best statistical precision and most stable fits in the presence of non-vanishing momentum transfer. The reason for this is that increasing $|\vpi|$ (or $|\vpf|$) causes a deterioration of the signal in $\tins$ close to $t_i$ ($t_f$), hence excluding as many noisy data points as possible improves the quality of the resulting signal. We remark that at least at vanishing momentum transfer ground state saturation can also be achieved by the plateau (or midpoint) method for sufficiently large values of $\tsep \gtrsim 2.5\fm$, as can be inferred from the effective form factors shown in the left column of Fig.~\ref{fig:eff_FF}. This is because there is no signal-to-noise problem for the pion at rest. Note that the excited state contamination in the full effective form factors including the quark-disconnected contribution is enhanced compared to the quark-connected part, i.e. the latter exhibits ground state saturation already for $\tsep\gtrsim1.5\fm$. Anyhow, the summation method is much less prone to fluctuations in the input data than the plateau method and it yields compatible but statistically more precise results.  In particular, it allows including data at smaller values of $\tsep$ because excited states are parametrically more strongly suppressed by $e^{-\Delta\tsep}$ compared to $e^{-\Delta\tsep/2}$ for e.g. the midpoint method. Finally, we remark that injecting one unit of momentum at the sink of the quark-connected piece generally leads to a statistically much more precise signal than injecting at the source, as expected. \par 

Besides the summation method, we have also attempted multi-state fits to extract the ground state matrix elements, explicitly fitting the time-dependence in $\tsep$ and $\tins$ simultaneously. However, we find that such fits suffer from severe stability issues due to the large covariance matrices and the very strong correlations in the ratio data. \par

\begin{figure}[t]
 \centering
 \includegraphics[totalheight=0.226\textheight]{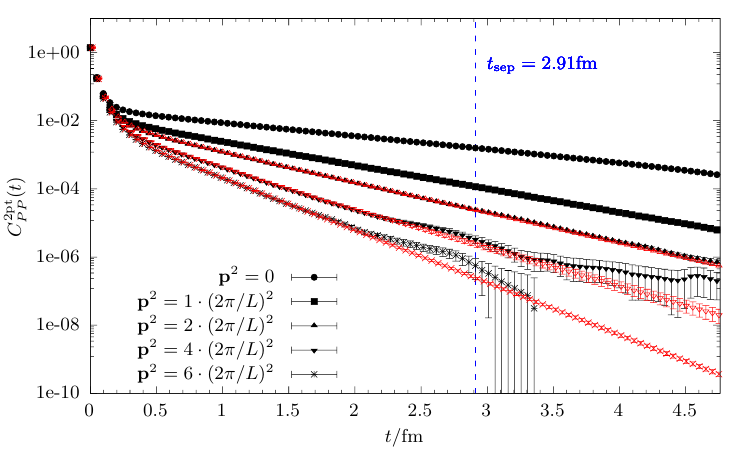}
 \includegraphics[totalheight=0.226\textheight]{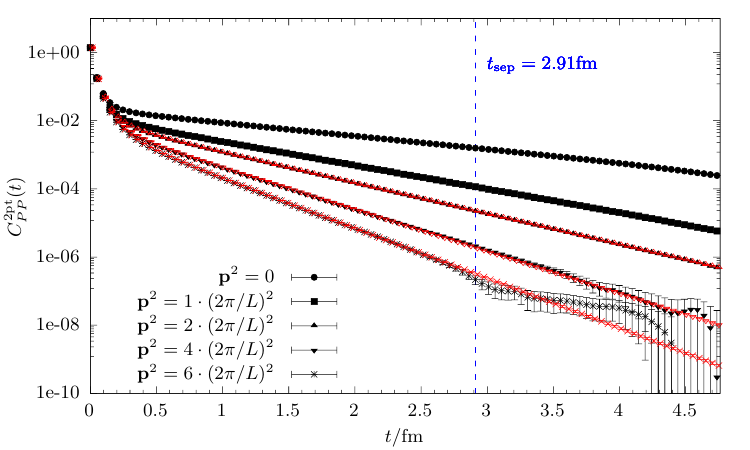}
 \caption{Signal for the two-point correlation function $\Ctwopt{PP}{\vp, t}$ (black symbols) on the E300 ensemble at several values of $\vp^2$. For $\vp^2\geq 2 \cdot (2\pi/L)^2$ the lattice data are replaced by the reconstructed signal (red symbols) that is obtained by replacing the large-$t$ tail with a fit of the ground state contribution. Results are shown for two-point statistics matching the quark-connected three-point function (left panel) and with quark-disconnected (2+1)-point function (right panel) at $\tsep=59a\approx2.91\fm$.}
 \label{fig:2pt_reconstruction}
\end{figure}

The emerging signal-to-noise problem at non vanishing values of $\vpi$ and $\vpf$ affects not only the three-point function but also the two-point functions that enter the ratio in Eq.~(\ref{eq:ratio}). This becomes an issue at larger values of $\tsep$ as $\Ctwopt{PP}{\vp, t}$ enters the ratio for any value $t\in[0,\tsep]$. An example is shown in Fig.~\ref{fig:2pt_reconstruction} for the E300 ensemble with two-point function statistics corresponding to the quark-connected and -disconnected three-point function at $\tsep=59a$. However, because $\Ctwopt{PP}{\vp, t}$ reaches ground-state saturation for $t\gtrsim 1.5\fm$ on every ensemble, the signal quality for the ratio can be improved by fitting the ground state of the two-point function and replacing its tail by the fit which is statistically much more precise than the available lattice data at large values of $t$. To this end, we fit $\Ctwopt{PP}{\vp, t}$ with a single state for $t\in\l[t_1,t_2\r]$, where $t_1$ is determined by an automatic scan on every ensemble for each value of $\vp^2$, and $t_2$ is chosen to be the last data point for which the signal is not yet lost in noise. The reconstructed signal is then obtained by replacing $\Ctwopt{PP}{\vp, t\geq t_1}$ with the fitted result on every bootstrap sample. While the momentum resolution can be very different depending on the ensemble, we find $\vp_\mathrm{min}^2=2\cdot(\pi/L)^2$ to be generally a reasonable choice for the smallest momentum from which on to employ this procedure. \par

 \section{Form factor parametrization} \label{sec:FF_parametrization}
\begin{figure}[t]
 \centering
 \includegraphics[totalheight=0.226\textheight]{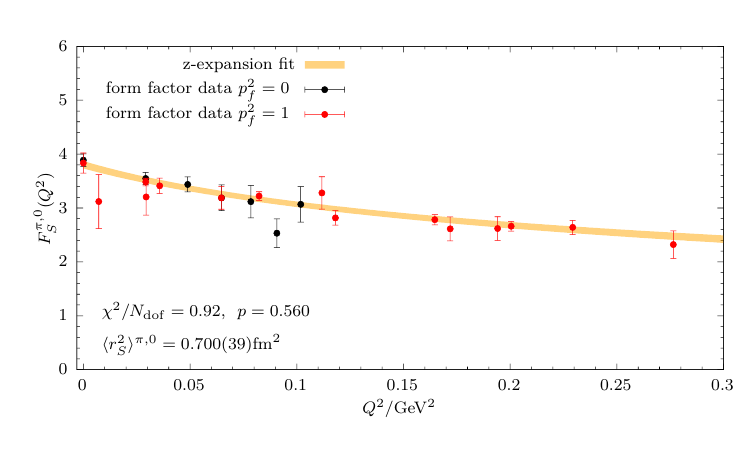}
 \includegraphics[totalheight=0.226\textheight]{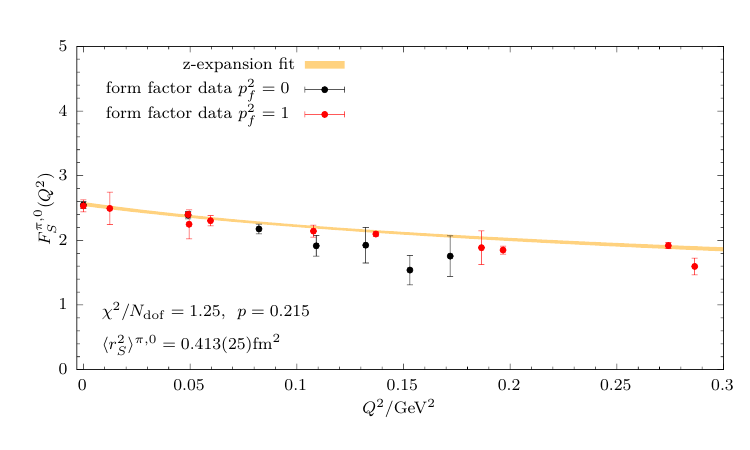}
 \includegraphics[totalheight=0.226\textheight]{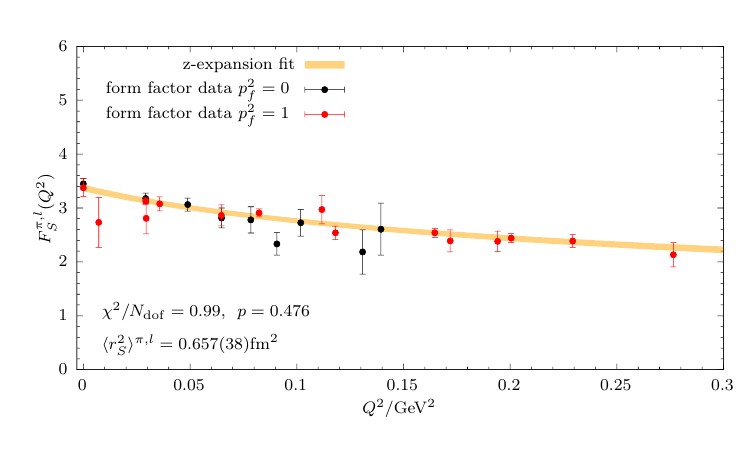}
 \includegraphics[totalheight=0.226\textheight]{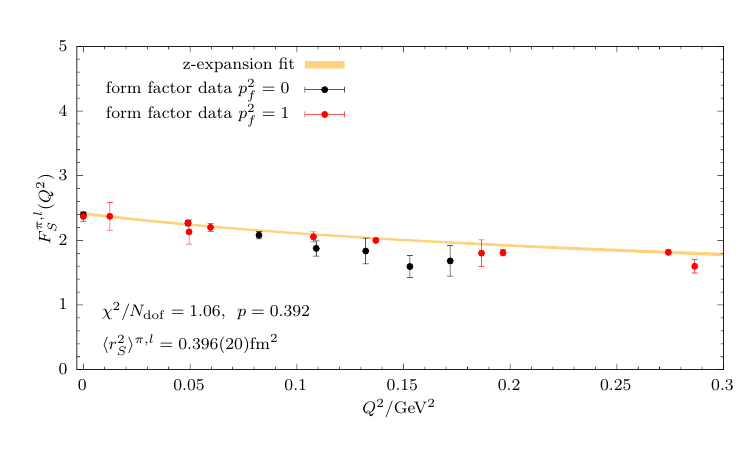}
 \includegraphics[totalheight=0.226\textheight]{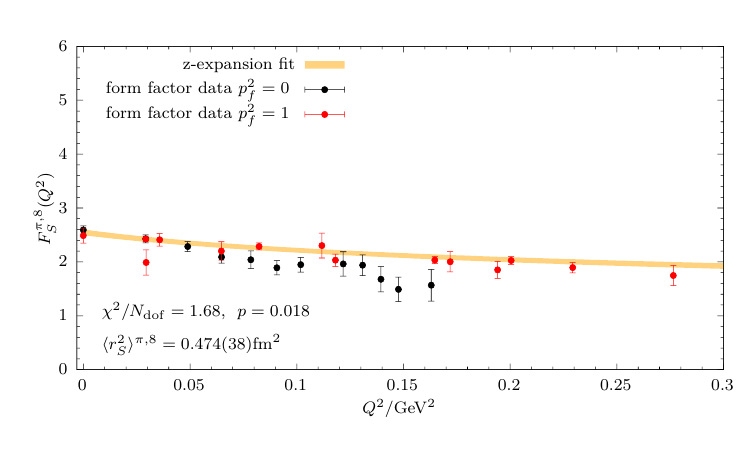}
 \includegraphics[totalheight=0.226\textheight]{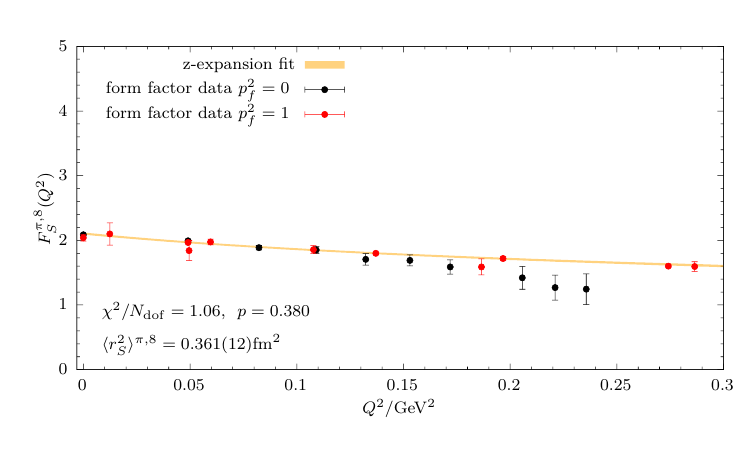}
 \caption{Form factor data and $z$-expansion fit bands for $\FFpisinglet{Q^2}$(top), $\FFpi{Q^2}$ (center) and $\FFpioctet{Q^2}$ (bottom) on the two most chiral ensembles E250 (left column) and E300 (right panel). Results are shown for the data set including source-sink separations with $1.25\fm\leq\tsep\leq3.25\fm$ and a cut in the momentum transfer of $Q^2\leq0.3\gev^2$.}
 \label{fig:FF_parametrization}
\end{figure}

The extraction of the scalar radius in Eq.~(\ref{eq:radius}) requires a parametrization of the lattice data for $F_S^{\pi,f}(Q^2)$ that is obtained at discrete values of $Q^2$. Previous lattice studies relied on rather crude linear interpolations between $F_S^{\pi,f}(0)$ and $F_S^{\pi,f}(Q^2)$ at the first non-vanishing momentum, as the required quark-disconnected contributions to $F_S^{\pi,f}(Q^2)$ could only be computed for one or two non-vanishing values of $Q^2$, see e.g. Refs.~\cite{Gulpers:2013uca,Gulpers:2015bba,Koponen:2015tkr}. Our present set of data allows us to reliably track $F_S^{\pi,f}(Q^2)$ for a much larger range of $Q^2$ values and with a fine-grained momentum resolution, especially on lattices with large physical volume as shown in Fig.~\ref{fig:FF_parametrization}. We observe that adding data with one unit of sink momentum greatly enhances the accessible range and resolution as well as the overall statistical quality of the signal. Therefore, we make use of a $z$-expansion fit ansatz~\cite{Hill:2010yb} to parametrize the $Q^2$-dependence
\begin{equation}
  F_S^{\pi,f}(Q^2) = \sum_{n=0}^{N_z} a_n^f z^n\,, \qquad z=\frac{\sqrt{t_\mathrm{cut}+Q^2} - \sqrt{\vphantom{Q^2}t_\mathrm{cut}-t_0}}{\sqrt{t_\mathrm{cut}+Q^2} + \sqrt{\vphantom{Q^2}t_\mathrm{cut}-t_0}} \,, 
 \label{eq:z_expansion}
\end{equation}
where we use $N_z=1$, $t_\mathrm{cut}=4M_\pi^2$ and $t_0 = t_0^\mathrm{opt} = t_\mathrm{cut}(1-\sqrt{1+Q^2_\mathrm{max}/t_\mathrm{cut}})$ \cite{Lee:2015jqa}. We find that increasing the order of the expansion $N_z \gg 1$ typically only inflates the error on the extracted radii $\< r_S^2\>_\pi^f \sim a_1^f$ without improving the quality of the fits. In particular for the most chiral ensembles E250 and E300 these fits yield an excellent description of the lattice data at the present level of statistics; cf. Fig~\ref{fig:FF_parametrization}. Anyhow, all fits are carried out for five different maximum values of the momentum transfer $Q^2_\mathrm{max}\in\l\{0.2,0.25,0.3,0.35,0.4\r\}\gev$ to account for residual higher-order effects in the final error budget. We remark, that there can be up to 125 distinct $Q^2$ values depending on the ensemble. Therefore, we have pruned the data that enter the fits (and Fig.~\ref{fig:FF_parametrization}) by demanding that the relative point errors do not exceed $20\%$ in case of $F_S^{\pi,0}(Q^2)$, $F_S^{\pi,l}(Q^2)$ and $10\%$ for $F_S^{\pi,8}(Q^2)$ which is inherently more precise. This leaves the fit results virtually unchanged as they are dominated by statistically much more precise data, but it prevents potential issues that may arise from estimating large and noisy covariance matrices in correlated fits. \par

 \section{Physical extrapolations} \label{sec:CCF}
For the physical extrapolations of the scalar radii we employ fit ans\"atze that are based on the corresponding NLO $SU(3)$ $\chi$PT expressions. They are obtained by rewriting Eqs.~(\ref{eq:rsqr0})--(\ref{eq:delta_rsqr_S}) in terms of leading order quark mass proxies $\xi_l=t_0 M_\pi^2$ and $\xi_s=t_0(2M_K^2-M_\pi^2)$, and expressing all dimensionful observables in units of $t_0$
\begin{alignat}{4}
 \frac{\rsqrpisinglet}{t_0} &= \frac{1}{(4\pi \sqrt{t_0} f_0)^2} \biggl[ 768\pi^2 \l(3L_4^r + L_5^r\r) \biggr.&- 19& + \frac{\xi_l}{\xi_l+2\xi_s}    &\l.- 12\log\l(\xi_l\r) - 6\log\l(\frac{\xi_l+\xi_s}{2}\r) \r] \,, \label{eq:r0_fit_model} \\
 \frac{\rsqrpi}{t_0}        &= \frac{1}{(4\pi \sqrt{t_0} f_0)^2} \biggl[ 768\pi^2 \l(2L_4^r + L_5^r\r) \biggr.&- 16& + \frac{\xi_l}{3(\xi_l+2\xi_s)} &\l.- 12\log\l(\xi_l\r) - 3\log\l(\frac{\xi_l+\xi_s}{2}\r) \r] \,, \label{eq:rl_fit_model} \\
 \frac{\rsqrpioctet}{t_0}   &= \frac{1}{(4\pi \sqrt{t_0} f_0)^2} \biggl[ 768\pi^2 L_5^r                \biggr.&- 10& - \frac{\xi_l}{\xi_l+2\xi_s}    &\l.- 12\log\l(\xi_l\r) + 3\log\l(\frac{\xi_l+\xi_s}{2}\r) \r] \,. \label{eq:r8_fit_model}
\end{alignat}
The scale-dependence $\mu$ has been absorbed into the definitions of $L_{4,5}^r$, implicitly setting $\sqrt{t_0} \mu = 1$ in the fits. In order to determine the physical values of the radii on any given set of input data, we individually fit the above expressions complemented by a term $\sim a^2/t_0$ with a free fit parameter to account for the continuum extrapolation. The physical point in the light and strange quark mass proxies is defined in the isospin limit using $M_\pi^\phys=134.8(3)\mev$ and $M_K^\phys=494.2(3)\mev$ from Ref.~\cite{Aoki:2016frl}. \par

In addition to fitting the full set of lattice data, we apply a set of data cuts to test for systematic effects in the physical extrapolations. First of all, we have implemented three different cuts in $M_\pi$, i.e. $M_\pi<M_\pi^\mathrm{cut}\in\l\{230, 265, 290\r\}\mev$ to test for residual higher-order effects in the light quark mass beyond the NLO expressions. Considering the steepness of the chiral extrapolation in the light quark mass due to the chiral logarithms in Eqs.~(\ref{eq:r0_fit_model})--(\ref{eq:r8_fit_model}) we expect this to be the most important source of uncertainty. Indeed, this is generally confirmed by our fits showing a clear trend towards better fit qualities for stricter $M_\pi$-cuts. Additionally, we apply a cut excluding the data at the coarsest lattice spacing $a<a^\mathrm{cut}=0.08\fm$, as well as a cut in the lattice volume $L<L^\mathrm{cut}=3.5\fm$. Moreover, the fits have been carried out for all possible combinations of these data cuts. Combined with the variations in the previous analysis steps, there is a total number of $N_M=910$ different models for the extrapolation of each of the three radii. In order to avoid fit stability issues that may arise for a few models with particularly small numbers of degrees of freedom $N_\mathrm{dof} \sim 4$ on some of the bootstrap samples, a prior has been put on $f_0$, i.e. the central value from the latest FLAG estimate~\cite{FlavourLatticeAveragingGroupFLAG:2021npn} $f_0^\mathrm{prior}=\sqrt{2} F_0^\mathrm{FLAG}=113.6\mev$ with a very broad width of 50\%. \par

\begin{figure}[t]
 \centering
 \includegraphics[totalheight=0.226\textheight]{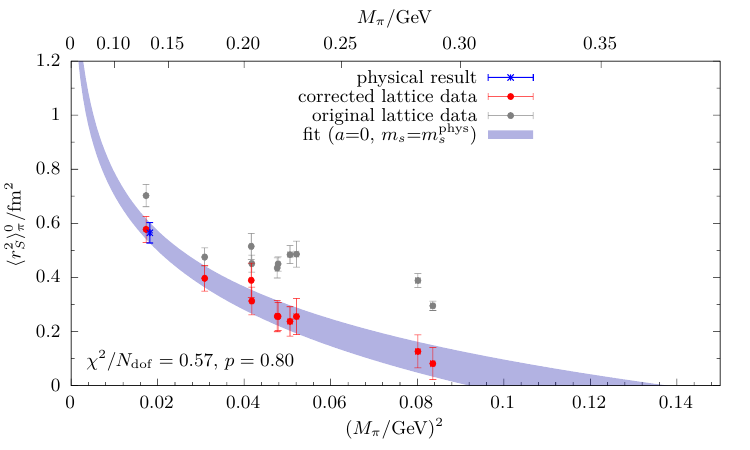}
 \includegraphics[totalheight=0.226\textheight]{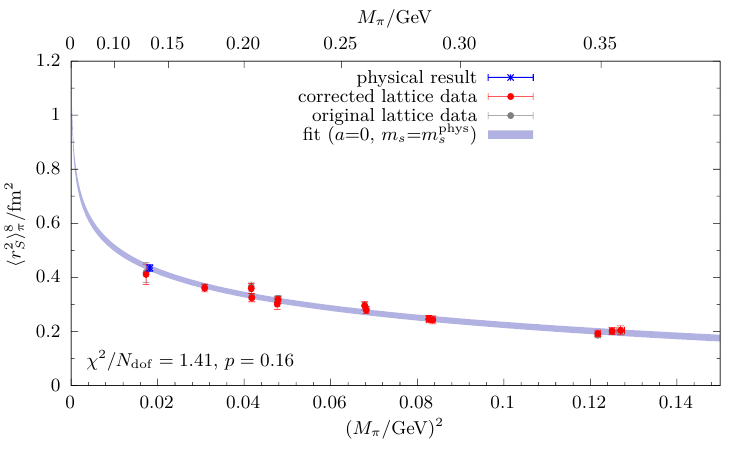} \\
 \includegraphics[totalheight=0.226\textheight]{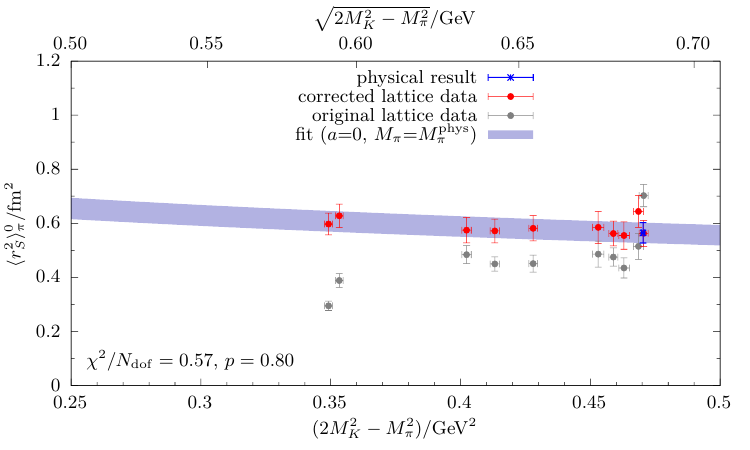}
 \includegraphics[totalheight=0.226\textheight]{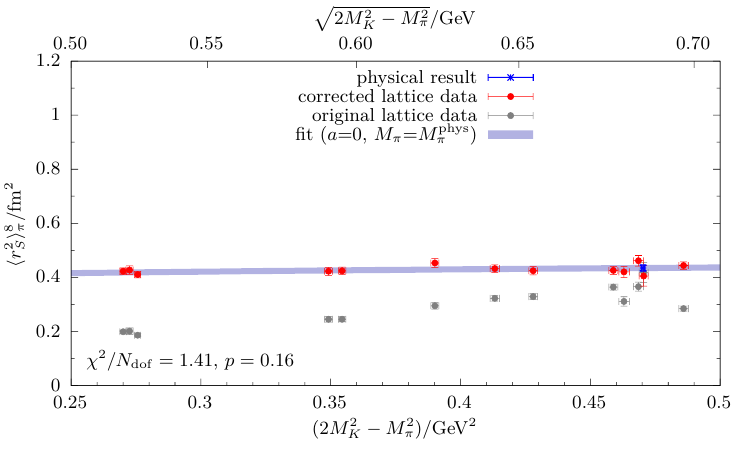} \\
 \includegraphics[totalheight=0.226\textheight]{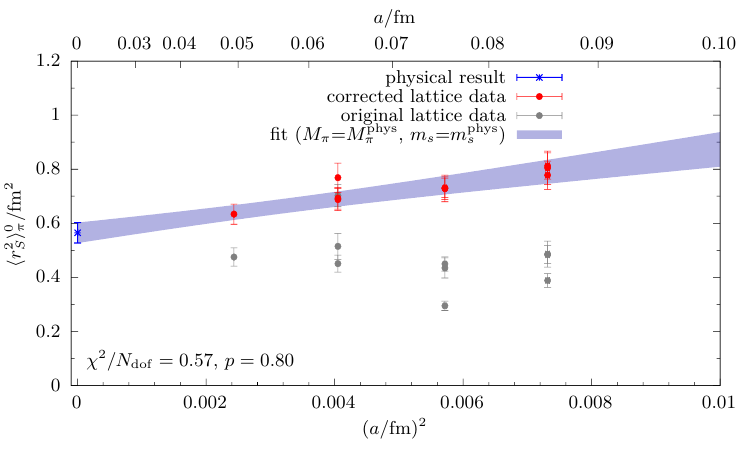}
 \includegraphics[totalheight=0.226\textheight]{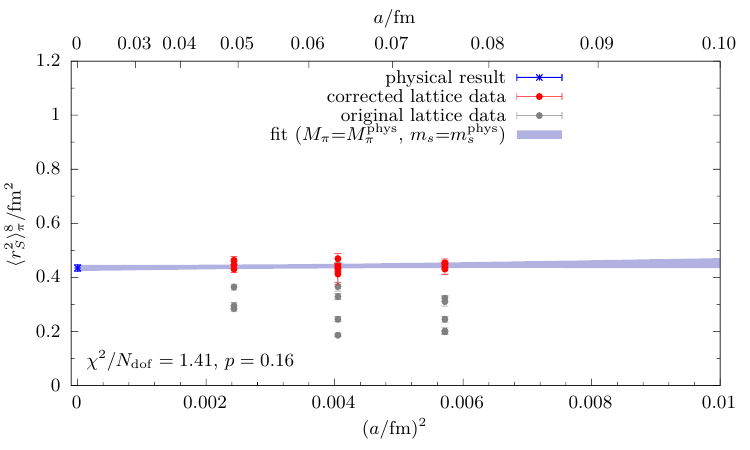}
 \caption{Examples for the physical extrapolation of $\rsqrpisinglet$ (left column) and $\rsqrpioctet$ (right column). The three figures for each observable show the extrapolation in the light and strange quark mass proxies $M_\pi^2$ and $2M_K^2-M_\pi^2$, as well as the continuum limit are obtained from the same fit. The red data have been corrected for the physical extrapolation in any other the variables that is not displayed on the horizontal axis in the respective panel, using the parameters obtained from the fit. Errors are statistical only.}
 \label{fig:CCF_octet_singlet_radius}
\end{figure}

\begin{figure}[t]
 \centering
  \includegraphics[totalheight=0.226\textheight]{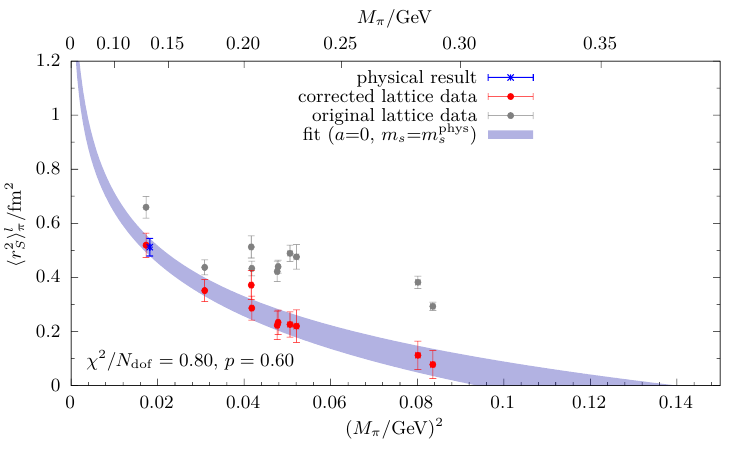}
  \includegraphics[totalheight=0.226\textheight]{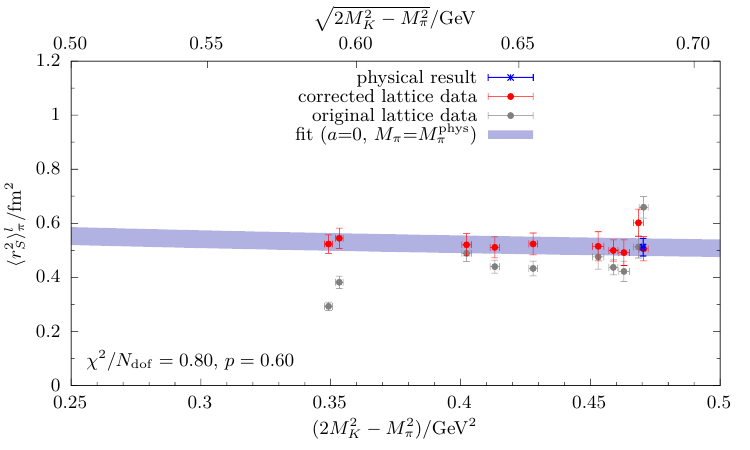} \\
  \includegraphics[totalheight=0.226\textheight]{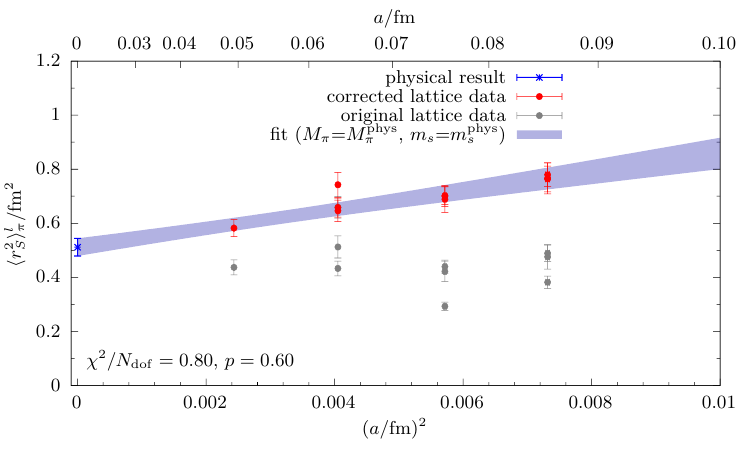} \\
 \caption{Same as Fig.~\ref{fig:CCF_octet_singlet_radius} but for $\rsqrpi$.}
 \label{fig:CCF_light_radius}
\end{figure}

Figure~\ref{fig:CCF_octet_singlet_radius} shows examples for the physical extrapolation of $\rsqrpisinglet$ and $\rsqrpioctet$ as a function of the light and strange quark mass proxies as well as $a^2$. Similarly, results for $\rsqrpi$ are shown in Fig.~\ref{fig:CCF_light_radius}. Note that the respective figures for $\rsqrpisinglet$, $\rsqrpioctet$ and $\rsqrpi$ are based on different input data sets (cuts) and have been selected to be representative of the final results for the radii obtained from the model average discussed in the next section. More specifically, the selected fits have been chosen such that they exhibit the largest weight while their physical result for the radius is within one standard deviation of the statistical error of the final result. \par

First of all, our lattice data clearly reproduces the expected slope in $M_\pi^2$ due to the presence of the corresponding chiral logs in Eqs.~(\ref{eq:r0_fit_model})--(\ref{eq:r8_fit_model}). In particular, this trend is confirmed by the most chiral ensemble E250 with slightly lighter-than-physical light quark mass. Corrections due to the continuum limit are generally non-negligible as well for $\rsqrpisinglet$ and $\rsqrpi$, whereas they tend to be smaller for $\rsqrpioctet$. However, we find that they can only be resolved reliably in the presence of data cuts in the pion mass. This can be taken as another hint that higher order effect become relevant for $M_\pi \gtrsim 250\mev$. A common feature of all three radii is that the chiral extrapolation in the strange quark mass is flat and essentially compatible with a constant. This feature remains very stable under data cuts. \par

Finally, concerning possible finite volume effects we have tested several approaches. Unlike data cuts in $M_\pi$ there is no clear picture for cuts in $L$ (or $a$). While a cut in $L$ can still have some effect when combined with other cuts, there is no systematic trend observed. Furthermore, there is no analytical expression for the finite size corrections to the pion scalar form factor available in the literature. However, we have tested the effect of adding corrections of the same form as the ones derived for meson masses and decay constants in Ref.~\cite{Colangelo:2005gd}, i.e. performing a simple substitution of the chiral logarithms
\begin{equation}
 \log\l(t_0 M_P^2\r) \rightarrow \log\l(t_0 M_P^2\r) + \tilde{g}_1(M_P L) \,, \quad P=\pi,K \,, 
 \label{eq:CDH_FV}
\end{equation}
where $\tilde{g}_1(M_P L)$ is the function defined in Eq.~(12) of Ref.~\cite{Colangelo:2005gd}. Somewhat surprisingly, this leads quite consistently to slightly worse $p$-values, whereas the resulting change to the physical values of the radii typically remains within the statistical uncertainty. Therefore, we refrain from including these variations into our final model averages. Furthermore, we have attempted to include a generic term of the form $\sim \tilde{g}_1(M_\pi L)$ with a free parameter as a coefficient. For the latter we find that our data is not statistically precise enough to reliably disentangle such a term. In particular, for models involving fewer input data due to the aforementioned cuts this often leads to unrealistically large, (anti-)correlated cancellations with other terms and unstable fits. At any rate, we conclude that there is no obvious finite-size trend in our data at the current level of precision. \par

% STATUS: Still need to reference discussion of data cuts in main text (?)

\begin{table}[!t]
 \caption{List of fit models used to extract each of the pertinent $SU(3)$ $\chi$PT LECs, i.e. linear combinations of the expressions given through Eqs.~(\ref{eq:r0_fit_model})--(\ref{eq:r8_fit_model}). The number of fit models $M_N$ also includes any further variations and (combinations of) data cuts at every stage of the analysis.}
 \centering
 \setlength{\tabcolsep}{0.5em}
 \begin{tabular}{llc}
  \hline\hline
  LEC   & fit models & $N_M$ \\
  \hline\hline
  $f_0$   & $\<r_S^2\>_\pi^0$,\ $\<r_S^2\>_\pi^l$,\ $\<r_S^2\>_\pi^8$,\ $\<r_S^2\>_\pi^0-\<r_S^2\>_\pi^l$,\ $\<r_S^2\>_\pi^0-\<r_S^2\>_\pi^8$,\ $\<r_S^2\>_\pi^l-\<r_S^2\>_\pi^8$ & 5460 \\
  $L_4^r$ & $\<r_S^2\>_\pi^0-\<r_S^2\>_\pi^l$,\ $\<r_S^2\>_\pi^0-\<r_S^2\>_\pi^8$,\ $\<r_S^2\>_\pi^l-\<r_S^2\>_\pi^8$ & 2730 \\
  $L_5^r$ & $\<r_S^2\>_\pi^8$,\ $3\<r_S^2\>_\pi^l-2\<r_S^2\>_\pi^0$ & 1820 \\
  \hline\hline
 \end{tabular}
 \label{tab:model_info}
\end{table}

For the determination of the individual LECs $L_4^r$ and $L_5^r$ it is convenient to fit suitable linear combinations of $\rsqrpisinglet$, $\rsqrpi$ and $\rsqrpioctet$ that give direct access to $L_4^r$ and $L_5^r$ as listed in Table~\ref{tab:model_info}. In particular $L_4^r$ is readily extracted e.g. from the difference $\rsqrpisinglet-\rsqrpi$ as it parametrizes a pure strange quark effect mediated by the quark-disconnected contribution. Again, a term $\sim a^2/t_0$ is added to every fit model to account for scaling artifacts. As for the individual radii, these fits are carried out for all cuts and variations of the input data as well, which is reflected by the number of models $N_M$ in Table~\ref{tab:model_info}. Note that the pion decay constant $f_0$ in the $SU(3)$ chiral limit is obtained from any such fit model, as it only appears in a global prefactor in Eqs.~(\ref{eq:r0_fit_model})--(\ref{eq:r8_fit_model}). We remark that we have also attempted simultaneous fits to the data for all three radii. While such fits inherently disentangle the contributions of individual LECs, they turn out to be problematic due to very strong correlations in the data resulting in a poorly estimated covariance matrix and typically unacceptable fit qualities. \par

 \section{Model averages and final results} \label{sec:AIC}
The individual results for the physical radii and LECs are combined in model averages to compute the final results with a full error budget. Taking into account the number of bootstrap samples there are $\mathcal{O}(10^6)$ fits per observable. For each of these observables, we assign weights \cite{doi:10.1177/0049124104268644,BMW:2014pzb,Neil:2022joj} to the respective models with index $n\in\{1,...,N_M\}$ and bootstrap sample $b\in\{1,...,N_B\}$,
\begin{equation}
  w_{n,b} = \frac{e^{-B_{n,b}}}{\sum_{k=1}^{N_M} e^{-B_{k,b}}} \,.
  \label{eq:weight}
\end{equation}
These weights are based on a variation of the Akaike information criterion (AIC) \cite{1100705,Akaike1998}, viz. the Bayesian AIC (BAIC) as introduced in Ref.~\cite{Neil:2022joj},
\begin{equation}
 B_{n,b} = \frac{1}{2}\chi^2_{n,b} + N_{\mathrm{par},n} + N_{\mathrm{cut},n} - N_{\mathrm{prio},n} \,,
 \label{eq:BAIC}
\end{equation}
where $\chi^2_{n,b}$ is the minimized, correlated $\chi^2$ obtained from the $n$-th fit model on the $b$-th bootstrap sample. $N_{\mathrm{par},n}$ and $N_{\mathrm{cut},n}$ denote the numbers of fit parameters and data removed by cuts for the respective model. Finally, the number of priors $N_{\mathrm{prio},n}$ is always the same, i.e. $N_{\mathrm{prio},n}=1$ because the fits do not require any priors besides the one on $f_0$ that is always present. For each observable $O$, we define the cumulative distribution function (CDF)
\begin{equation}
 CDF(y)= \frac{1}{N_B} \sum_{n=1}^{N_M} \sum_{b=1}^{N_B} w_{n,b} \Theta(y-O_{n,b}) \,,
 \label{eq:CDF}
\end{equation}
where $\Theta$ is the Heaviside step function. Note that the CDFs are constructed in a fully non-parametric way, i.e. by directly using the bootstrap distributions which differs from the approach discussed in Ref.~\cite{Borsanyi:2020mff}, that relies on (weighted) Gaussian CDFs derived only from the central values and errors of the individual models. The median and the quantiles that correspond to $1\sigma$ errors for a Gaussian distribution define the central value and total error of the final results, respectively. Moreover, the total error is subdivided into a statistical and systematic contribution by a procedure similar to the one introduced in Ref.~\cite{Borsanyi:2020mff}. The only required modification is that the direct rescaling of the statistical errors has to be replaced by a rescaling of the widths of the bootstrap distributions in our approach. For the rescaling factor we choose $\lambda=2$. \par

\begin{figure}[t]
 \centering
  \includegraphics[totalheight=0.226\textheight]{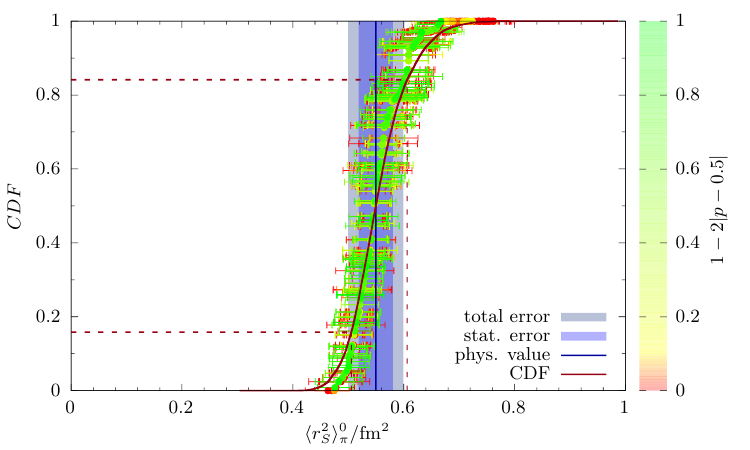}
  \includegraphics[totalheight=0.226\textheight]{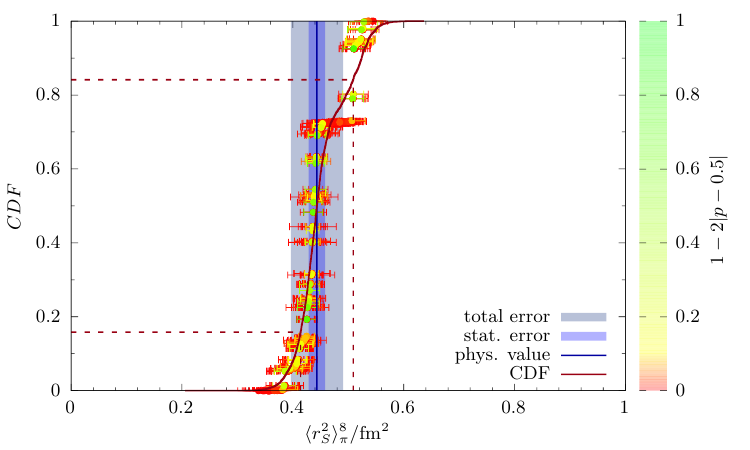}
  \caption{Cumulative distribution functions (CDFs) of the results for the singlet and octet scalar radii $\rsqrpisinglet$ and $\rsqrpioctet$. Each data point represents the result and statistical error from an individual fit model and its color is determined by the corresponding p-value weight indicating the quality of the fit. Note that the later is different from the Akaike weight actually used to obtain the CDF. The solid blue line indicates the final result from the model average and together with its statistical and full error bands. The dashed lines correspond to the $1\sigma$-quantiles of the CDF.}
  \label{fig:CDF_rsqr_0_and_8}
\end{figure}

\begin{figure}[t]
 \centering
  \includegraphics[totalheight=0.226\textheight]{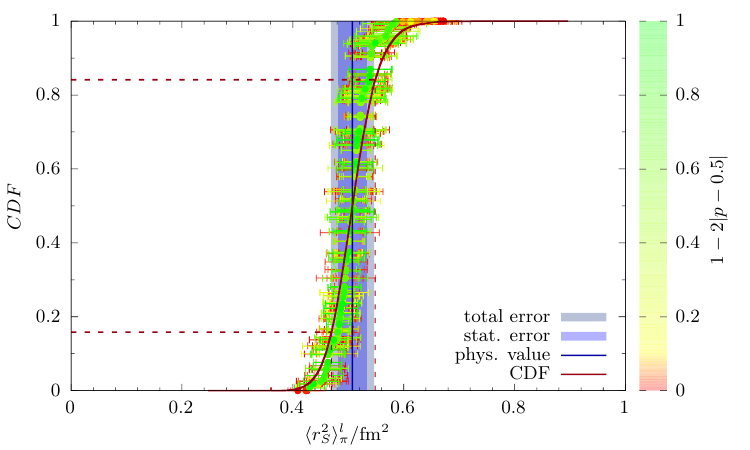}
  \includegraphics[totalheight=0.226\textheight]{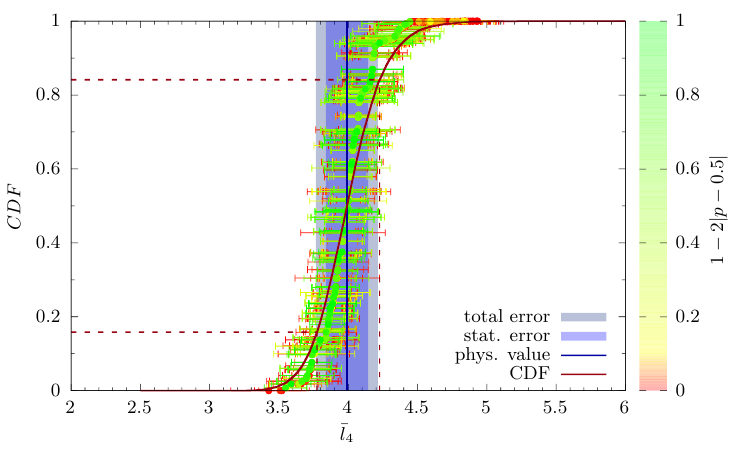} 
  \caption{Same as Fig.~\ref{fig:CDF_rsqr_l_and_l4bar} but for $\rsqrpi$ and the closely related LEC $\bar{\ell}_4$.}
 \label{fig:CDF_rsqr_l_and_l4bar}
\end{figure}

Figure~\ref{fig:CDF_rsqr_0_and_8} shows the CDFs and results for $\rsqrpisinglet$ and $\rsqrpioctet$. The CDF for $\rsqrpisinglet$ is in agreement with the assumption of Gaussian-distributed model data, whereas some skewness is observed for $\rsqrpioctet$. However, the data for the latter is statistically much more precise due to the (relative) smallness and precision of the $l-s$ quark-disconnected contribution, which is reflected by worse fit qualities for the individual models and results in an increased systematic error. The CDF for $\rsqrpi$ is shown in the left panel of Fig.~\ref{fig:CDF_rsqr_l_and_l4bar} and is fairly similar to the one for $\rsqrpisinglet$ concerning its shape and overall fit qualities of the underlying data, as expected. Our final results for the physical values of the three radii read
\begin{align}
 \rsqrpisinglet &= 0.550 \staterr{32} \syserr{39} \totalerr{50} \fm^2 \,, \label{eq:rsqr_0_phys} \\
 \rsqrpi        &= 0.508 \staterr{27} \syserr{29} \totalerr{39} \fm^2 \,, \label{eq:rsqr_l_phys} \\
 \rsqrpioctet   &= 0.443 \staterr{16} \syserr{45} \totalerr{47} \fm^2 \,, \label{eq:rsqr_8_phys}
\end{align}
where the systematic error comprises a full error budget, i.e. the residual uncertainty due to excited states, form factor parametrization and the physical-point extrapolation. Comparing these results to the only other available LQCD calculation involving strange quarks in Ref.~\cite{Koponen:2015tkr}, we find that our results are larger but in individual agreement within errors. The two-flavor calculation in Refs.~\cite{Gulpers:2013uca,Gulpers:2015bba}, on the other hand, predicts a value for $\rsqrpi$ that is about $2\sigma$ larger than ours. This could readily be explained by the effect of the strange quark contribution, but we stress that our present calculation also exhibits much better control over various systematic effects. \par

The $\bar{\ell}_4$ LEC in SU(2) \chiPT{} is directly related to $\rsqrpi$ via Eq.~(\ref{eq:rsqrl_SU2}), which leads to a fairly similar shape for its CDF in the right panel of Fig.~\ref{fig:CDF_rsqr_l_and_l4bar}. The final result is given by
\begin{equation}
 \bar{\ell}_4 = 3.99\staterr{15}\syserr{17}\totalerr{23} \,.
 \label{eq:l4bar_phys}
\end{equation}
where we have used $f_{\pi,\phys}=130.2(1.2)\mev$ from Ref.~\cite{Zyla:2020zbs} as input to evaluate the r.h.s. of Eq.~(\ref{eq:rsqrl_SU2}). The value is in excellent agreement with the most recent FLAG estimate~\cite{FlavourLatticeAveragingGroupFLAG:2021npn} $\bar{\ell}_4^\mathrm{FLAG} = 4.02(45)$ for $N_f=2+1$ flavors, but with only half the total error. Besides, our result is the first LQCD determination of $\bar{\ell}_4$ with $N_f=2+1$ flavors from a form factor calculation, whereas all previous LQCD results~\cite{MILC:2010hzw,Beane:2011zm,Borsanyi:2012zv,BMW:2013fzj,Boyle:2015exm} that enter the FLAG average are derived from decay constant calculations based on meson two-point functions. Finally, our result is in broad agreement with phenomenological analyses in e.g. Refs.~\cite{Gasser:1983yg,Bijnens:1998fm,Amoros:2000mc,Colangelo:2001df}, although their results show a trend towards $\bar{\ell}_4>4$. \par

\begin{figure}[t]
 \centering
 \includegraphics[totalheight=0.226\textheight]{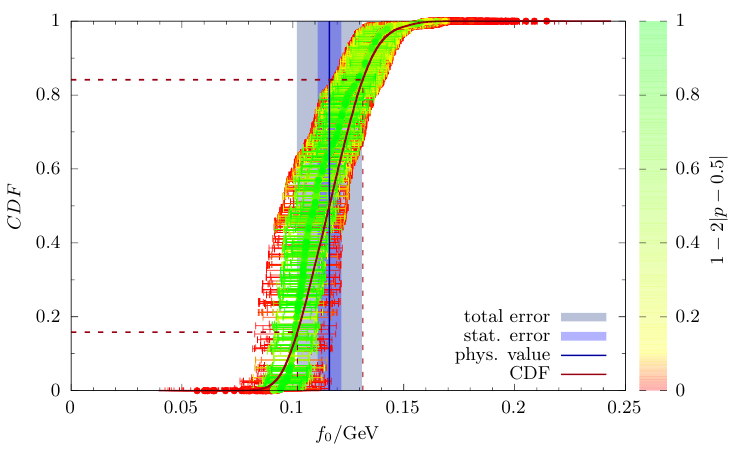}
 \includegraphics[totalheight=0.226\textheight]{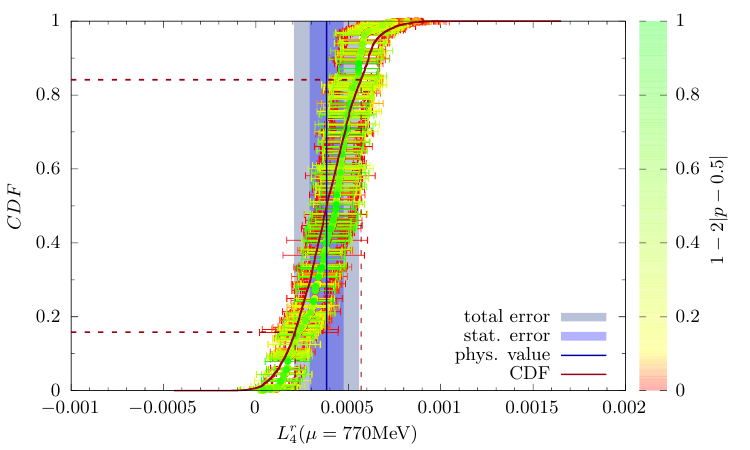} \\
 \includegraphics[totalheight=0.226\textheight]{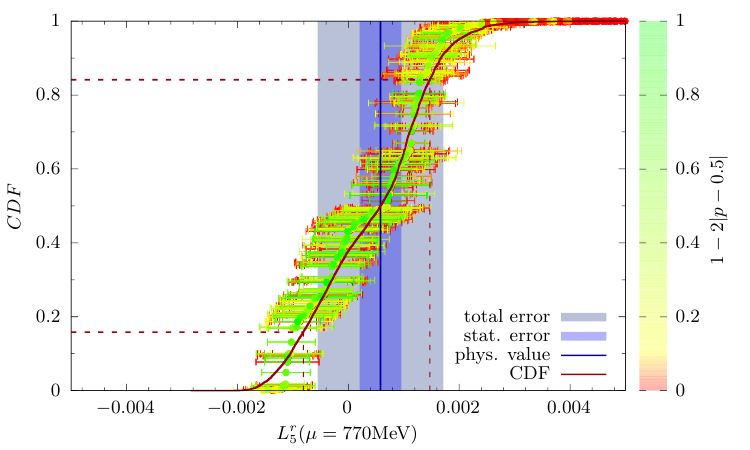}
 \caption{Same as Fig.~\ref{fig:CDF_rsqr_l_and_l4bar} but for the $\SU{3}$ LECs $f_0$, $L_4^r(\mu)$ and $L_5^r(\mu)$ at a scale of $\mu=770\mev$.}
 \label{fig:CDF_SU3_LECs}
\end{figure}

For the SU(3) LECs $f_0$, $L_4^r$ and $L_5^r$ the final results are based on the models in Table~\ref{tab:model_info}, which is reflected by the larger number of data points in Fig.~\ref{fig:CDF_SU3_LECs}. For the pion decay constant in the SU(3) chiral limit we find
\begin{equation}
 f_0 = 116.5\staterr{5.5}\syserr{13.7}\totalerr{14.8} \mev \,. \label{eq:f_0_phys}
\end{equation}
which agrees with the FLAG value~\cite{FlavourLatticeAveragingGroupFLAG:2021npn,MILC:2010hzw} $f_0 = 114.0(8.5)\mev$, whereas a more recent study by the $\chi$QCD collaboration in Ref.~\cite{Liang:2021pql} prefers a much smaller value of $f_0=82.3(0.7)(14.1)\mev$ in stark contrast to our result. On the other hand, our value for $f_0$ agrees almost exactly with the semiphenomenological result $f_0=116.46(96)\mev$ in Ref.~\cite{Lutz:2024ubv} from a N$^3$LO global fit of octet and decuplet baryon masses on the CLS ensembles. \par

The final results for the NLO LECs have been evaluated at a scale of $\mu=770\mev$
\begin{align}
 L_4^r(\mu=770\mev) &= +0.38\staterr{09}\syserr{15}\totalerr{18} \times 10^{-3}       \,, \label{eq:L_4_phys} \\
 L_5^r(\mu=770\mev) &= +0.58\staterr{0.38}\syserr{1.08}\totalerr{1.14} \times 10^{-3} \,. \label{eq:L_5_phys}
\end{align}
Our determination for $L_4^r$ is the first with controlled systematics and sufficient statistical precision to obtain a non-zero value. Furthermore, it is the only direct determination based on a form factor calculation. The only other two lattice results $L_4^r(\mu=770\mev) = -0.02(56) \times 10^{-3}$ ($N_f=2+1$, from Ref.~\cite{MILC:2010hzw}) and $L_4^r(\mu=770\mev)=+0.09(34)$ ($N_f=2+1+1$, from Ref.~\cite{Dowdall:2013rya}) are compatible within errors but do not even agree on a common sign. Unlike for $L_4^r$ we find that our fits are not particularly sensitive to $L_5^r$, resulting in a rather substantial model spread in the lower panel of Fig.~\ref{fig:CDF_SU3_LECs}. This results in a significantly increased total error that is dominated by systematic uncertainty. However, the central value has the correct sign and roughly same magnitude as other lattice determination from decay constant fits; cf. Refs.~\cite{MILC:2010hzw,Dowdall:2013rya}.

 \section{Summary and outlook} \label{sec:summary}
In this study we have computed physical results for the pion scalar radii $\rsqrpisinglet$, $\rsqrpioctet$ and $\rsqrpi$, as well as the pertinent LECs in the respective NLO expressions in SU(2) and SU(3) \chiPT{}. The results feature a full error budget covering the systematic uncertainties at every stage of the analysis. This was made possible mainly by calculating the quark-disconnected contributions to $F^{\pi,f}_S(Q^2)$ with unprecedented statistical precision and momentum resolution. For the first time, this enabled us to e.g. extract the pion scalar radii from a parametrization of the form factors beyond a simple, linear interpolation, and to obtain a non-zero result for $L_4^r$ that depends entirely on the quark-disconnected contribution of the strange quark. Besides, our calculation confirms the importance of the quark-disconnected contribution for the pion scalar radius as predicted in Ref.~\cite{Juttner:2012xs} from a \chiPT{} analysis. Furthermore, the statistical precision of the lattice data is mirrored by our result for the SU(2) LEC $\bar{\ell}_4$, that crucially depends on the light quark-disconnected contribution to the form factor. Lastly, our calculation confirms the observation in Refs.~\cite{Gulpers:2015bba,Koponen:2015tkr} that the quark-disconnected contribution to $F^{\pi,f}_S(Q^2)$ is affected by large lattice artifacts in the case of Wilson fermions. \par

In the future, it might be interesting to include additional ensembles at physical quark mass and even finer lattice spacing to further reduce systematic errors. Besides, since we have also computed and stored all local and one-link displaced insertion operators for the quark-connected three-point function, our calculation could rather easily be extended to further observables, such as e.g. the pion and kaon vector radii and average quark momentum fractions. \par

 \section*{Acknowledgments}
We thank Stephan D\"urr and Andreas J\"uttner for useful discussions. This research is supported by the Deutsche Forschungsgemeinschaft (DFG, German Research Foundation) through project HI~2048/1-3 (project No.~399400745). The authors gratefully acknowledge the Gauss Centre for Supercomputing e.V. (www.gauss-centre.eu) for funding this project by providing computing time on the GCS Supercomputer SuperMUC-NG at Leibniz Supercomputing Centre and on the GCS Supercomputers JUQUEEN\cite{juqueen} and JUWELS\cite{JUWELS} at Jülich Supercomputing Centre (JSC). The authors gratefully acknowledge the computing time made available to them on the high-performance computer Mogon-NHR at the NHR Centre NHR S\"ud-West. This center is jointly supported by the Federal Ministry of Education and Research and the state governments participating in the NHR (www.nhr-verein.de/unsere-partner). Additional calculations have been performed on the HPC clusters Clover at the Helmholtz-Institut Mainz, and Mogon II and HIMster-2 at Johannes-Gutenberg Universit\"at Mainz. The QDP++ library \cite{Edwards:2004sx} and the deflated SAP+GCR solver from the openQCD package \cite{openQCD} have been used in our simulation code. We thank our colleagues in the CLS initiative for sharing gauge ensembles.

 \bibliography{refs}
 \clearpage

\end{document}